\documentclass[11pt]{article}

\usepackage{times}
\usepackage{helvet}
\usepackage{courier}

\usepackage{graphicx,pstricks}
\usepackage{moreverb}
\usepackage{subfigure}
\usepackage{epsfig}
\usepackage{subfigure}
\usepackage{setspace}

\usepackage{fullpage}
\usepackage{enumerate}
\usepackage{amsfonts}
\usepackage{amssymb}
\usepackage{amsthm}
\usepackage{graphicx}
\usepackage{amsmath}
\usepackage{array}

\newtheorem{theorem}{Theorem}[section]
\newtheorem{lemma}[theorem]{Lemma}

\newtheorem{definition}[theorem]{Definition}

\newcommand{\rtoldownimplies}{ \rotatebox[origin=c]{45}{$\Leftarrow$}}

\newcommand{\subseteqdown}{ \rotatebox[origin=c]{90}{$\supseteq$}}
\newcommand{\dotsdown}{ \rotatebox[origin=c]{90}{$\cdots$}}
\newcommand{\dotsdiag}{ \rotatebox[origin=c]{45}{$\cdots$}}

\begin{document}

\title{
\textbf{Maximizing the Spread of Stable Influence:\\
 Leveraging Norm-driven
Moral-Motivation \\
for Green Behavior Change in Networks}}


\date{}

\author{Gwen Spencer\thanks{Neukom Postdoctoral Fellow, Neukom Institute for Computational Science, Dartmouth College.}   and Richard B. Howarth\thanks{Dartmouth College.}}

\maketitle

\begin{abstract}
In an effort to understand why individuals choose to participate in personally-expensive pro-environmental behaviors,
environmental and behavioral economists have examined a moral-motivation model in which the decision to adopt a pro-environmental behavior depends
on the society-wide market share of that behavior.
An increasing body of practical research on adoption of pro-environmental behavior emphasizes the importance of
encouragement from local social contacts and messaging about locally-embraced norms: we respond by
extending the moral-motivation model to a social networks setting.
We obtain a new decision rule: an individual adopts a pro-environmental behavior if he or she observes a certain threshold of adoption within their local
social neighborhood. This gives rise to a concurrent update process which describes adoption of a pro-environmental behavior  spreading through a network. The process evolves according to a set of difference equations in an exponentially large space that describes all possible patterns of adoption. The original moral-motivation model corresponds to the special case of our network version in a complete graph.

In parallel with the original moral-motivation work, we examine issues of stability of adoption (which now depends intimately on
the spatial distribution of adoption within the network), bounds on the convergence time of the process, and the implications of
the model on potential impacts of periods of temporary subsidy.  In particular, we are interested in the planning question of how periods of temporary subsidy may be targeted in the network in order to ensure migration to a stable equilibrium with a high rate of participation
in the pro-environmental behavior.  At such a \textit{green equilibrium}, stability is enforced by egoistic utility benefits individuals experience from conforming to a locally-embraced green norm (and where positive externalities associated with high environment quality accrue to the entire society). To examine this issue, we create a time-indexed Integer Program Model (which has modest size due to new general theorems we prove on convergence times) that allows practically-efficient measurement of the exact optimal set of individuals to target with subsidy. Making a connection to the theoretical computer science literature, we prove that no rigorously-efficient method exists to compute such a set. Our results are general and do not require that the network have any specific topological properties.

The qualitative predictions of the network moral-motivation model can depart strongly from the predictions of the original moral-motivation model: for certain classes of networks, careful targeting of subsidy within a network can greatly reduce the number of subsidized individuals required to reach the green equilibrium. Asymptotic classes of examples demonstrate that such departures can be severe. Further, computational experiments show substantial departures for networks that resemble real social networks: we deploy our IP model with off-the-shelf solvers in modestly-sized highly-clustered small-world networks related to the famous small-world networks of Watts and Strogatz.
\end{abstract}

\section{Introduction}

\textbf{Moral-motivation in Environmental and Behavioral Economics.}
Voluntary participation in personally-expensive pro-environmental
behaviors (like opting-in to pay a higher rate for electricity that is
partially supplied by renewable sources like solar and wind) appears illogical
under classic economic models. Nevertheless, this option
is offered by electric utilities in many parts of the United States, and some members of the
public opt to pay the higher rates \cite{Nyborg2006351}.
While the adoption of some pro-environmental behaviors
(like curbside recycling) may be actively
encouraged by peer-observation
(and corresponding fear of social dis-utility
for not conforming-a phenomenon studied under the term ``social sanctioning''), adoption of
minimally-observable pro-environmental
behaviors led environmental and behavioral economists to
explore models that include utility derived
from ``moral motivation.'' See \cite{Nyborg2006351}, \cite{Proenvironmentalbehaviorrationalchoicemeets}. These models
posit that individuals experience
utility when they feel that they
are acting in accordance with a virtuous
norm.  In particular,
the moral-motivation model proposes that
this utility benefit is non-decreasing in the percentage of
adoption society-wide.

Colloquially, we might describe the moral-motivation model as follows:
if an individual perceives a personally-expensive (in terms
of time, money, etc.) green behavior to be a normal part of good citizenship,
then she can effectively be compensated for spending money or time engaging
in the green behavior by the satisfaction of being a good citizen.
A natural policy question then arises \cite{Nyborg2006351}: for settings where moral-motivation applies,
how can we best motivate many individuals to adopt green behavior?\\

\noindent \textbf{The Importance of a Local View.} In \cite{Nyborg2006351}, Nyborg, Howarth and Brekke relax the assumption
that individuals have perfect knowledge
of society-wide adoption rates,
but individuals remain essentially symmetric: knowledge is impacted
by stochastic noise, but every member of the society
sees the same noisy picture of society-wide adoption rates.
In strong contrast, recent studies on attitudes and
adoption of emerging green-technologies and green behavior highlight the
importance of normative encouragement from local social contacts and messaging about locally-embraced norms of behavior.

In a qualitative study on consumer attitudes from the Institute of Transportation Studies at U.C. Davis, entitled ``Interpersonal influence within car buyer's social networks: applying five perspectives to plug-in hybrid vehicle drivers," Axsen and Kurani \cite{Axsen_Kurani_2009} examine several frameworks that aim to understand how positive views of plug-in hybrid electric vehicles (PHEVs) are shaped.  Eleven families test-drove PHEVs for a 4-6 week
period; their perceptions of the usefulness and convenience of  the technology and their interactions with social contacts were tracked. Axsen and Kurani highlight ``the existence of supportive pro-societal values within the household's
social network," as one of three key factors that contribute to a high valuation of the benefits of PHEVs at the end of the trial. This factor is important
even when individuals have a technical understanding of the environmental benefits of PHEVs. Axsen and Kurani conclude,
\begin{center}
\begin{minipage}{12cm}
\textit{``...integrating concepts from contagion, conformity, and reflexivity...this paper points to the importance of: disseminating functional awareness of such technologies, stimulating interpersonal discussion of prosocietal benefits, and \textbf{marketing to a social network rather than only the individual} car buyer."} \cite{Axsen_Kurani_2009}
\end{minipage}
\end{center}

Geographic proximity may also contribute to
the local (and thus, spatially heterogeneous) nature of norms about green behavior participation.  In a study from the Journal of
Consumer Research, entitled, ``A Room with a Viewpoint: Using Social Norms to Motivate Environmental Conservation in Hotels,"  Goldstein, Cialdini and Griskevicius \cite{roomwithview} conduct a sizable controlled experiment on how different norm-based messages impact the reuse of towels by hotel guests.  Switching from a standard door-hanger message about saving the environment to a message about other hotel guests reusing towels at a high rate resulted in a $26\%$ jump in towel reuse. Replacing a message of the later type with a message that other hotel guests \textit{in the same room} reused towels at a high rate resulted in a further $15\%$ jump in towel reuse.  Goldstein, et al. describe their observations as the activation of a desire to conform to \textit{provincial norms}: they warn that attempts to market pro-environmental behaviors by appealing to broader demographic categories (e.g. to women) may waste valuable opportunities.

In describing the ``environmental morale," of individuals, Frey and Stutzer \cite{Environmentalmorale} summarize the ``core results" of laboratory experiments on when individuals will provide for the common good despite personal costs. Three of seven such results emphasize the importance of repeated contact with the same people (including face-to-face contact) in determining when individuals will be willing to cooperate.\\

\noindent \textbf{Our Extension: Network Moral Motivation.}  Strong empirical evidence suggests that
people infer prevailing pro-environmental norms based on the behavior of the people they encounter and engage with, and these norms are adopted in response to both internal motivation and social pressure. We respond by extending the moral-motivation model from \cite{Nyborg2006351} to networks in which individual behavior depends on the
percentage of adoption among neighbors in the network. Each individual in the network has a threshold: if the percentage of
their neighbors that adopt a green behavior is above the threshold, they also adopt the green behavior (this decision rule corresponds to the egoistic benefit of conforming to a virtuous green norm overcoming the other costs associated with that behavior).
Our work contributes to the growing exploration in the economic literature of how studied models
may be extended to network settings (for example, the extensions of many game-theoretic notions
to networks, as in \cite{galeotti2010network} and \cite{whoswhoWantedthekeyplayer}, the latter of which specifically focuses on evaluating how each player-based on network position-impacts the equilibrium outcome).\footnote{With different connotations, economists have used the term \textit{Network effect} to describe systems in which
positive externalities accrue to members of a network as a set of adopters grows: our model
exhibits this property locally within the social network.}

The decision rule described above gives rise to green behavior spreading
through the network via a spread mechanism highly similar to the
recently-studied notion of \textit{complex contagion} spreading through a network in the mathematical sociology literature (see the work of Centola et al. \cite{Cascadedynamicsofcomplexcontagion},\cite{Complexcontagionsandtheweaknessoflongties}).  The main modeling difference is that, following the extensive body of work on extending game theory to networks, our model allows each individual to update their adoption status in every time step (decisions to update are concurrent). We also focus on different questions.
The work in sociology uses simulation to ask questions
about what network properties
best propagate a new behavior that arises in a random neighborhood of the network.
In contrast, we are interested in planning questions in which a specific social
network exists: how can knowledge of this network inform the design of targeted campaigns to spread and normalize pro-environmental behaviors?

The (non-network) moral-motivation model creates a framework
with important implications for policies that aim to
encourage pro-environmental behavior.
In particular, the findings in \cite{Nyborg2006351} predict that
temporary periods of financial subsidy can cause
migration from an equilibrium in which
no members of the society adopt the pro-environmental
behavior to an equilibrium in which a high percentage
of the society adopts the behavior.
Once this favorable equilibrium is reached, the society accrues
additional environmental benefits associated with widespread pro-environmental behavior.
The idea is that
a temporary financial subsidy is implemented (so that adopting
the pro-environmental behavior gives a utility gain regardless of
adoption rates) until some critical level of society-wide adoption is
reached.  Once the critical
level of adoption is reached, the egoistic utility benefit of conforming
to the widely-adopted pro-environmental norm is sufficient to guarantee convergence
to a high-adoption equilibrium.  Without
the adoption-percentage-driven moral-motivation benefit, classical
models predict that when temporary subsidies are removed,
participation in the pro-environmental behavior will disappear.

In parallel with this original work on moral-motivation, we focus on
on the policy/planning question in networks where individuals have access to adoption information only among their neighbors.
How can understanding moral-motivation in networks allow planners to leverage
the boom in emerging information about social networks to
create temporary targeted subsidies, or targeted public education campaigns, that normalize pro-environmental behavior?
As in the non-network moral-motivation model the goal is to convert the society to a stable ``virtuous equilibrium" characterized by high levels of pro-environmental behavior.
Stability
of this equilibrium after the temporary subsidy lapses again must be enforced
by the moral-motivation of individuals, however, in the network setting the
issue of stability has become much more complex. Suppose that a particular member of the network is an
adopter when the subsidy is in place, but when it lapses, an insufficient number of their neighbors are adopters
to convince them to adopt due to moral-motivation alone: when they switch to non-adoption, each of their neighbors will also
recompute whether participating in the green behavior gives positive net utility, etc.
Effectively, at the end of an intervention period, participation in pro-environmental behavior may be eroded by cascades that domino through the network. The potential for such erosion has been described qualitatively by those interested
in persistence of sustainability-oriented policies:

\begin{center}
\begin{minipage}{12cm}
\textit{``Regulations and economic incentives play an important role in encouraging changes in behavior, but although these may change practices.... Without changes in social norms, people often revert to the old ways when incentives end or regulations are no longer enforced, and so long-term protection may be compromised."-Jules Pretty, 2003, Science} \cite{Pretty12122003}
\end{minipage}
\end{center}

To choose a good set of individuals to temporarily subsidize, we must characterize the long-term effect of subsidy.  This effect depends on a spatially-heterogeneous pattern of adoption that evolves according to the local update rules we have described above. A technical contribution of this work is proving that
the evolving pattern of adoption converges (almost), and rather quickly.
There is one notion of
temporary subsidy for which the evolution of the process can be cleanly divided into a phase in which adoption is growing and
a second phase in which it is eroding. In this setting it is straightforward to prove a bound on convergence time that is linear in the number of individuals in the network.
For another natural notion of temporary subsidy in which adoption and erosion of adoption proceed simultaneously, the situation is more complicated.
Here we extend results of Goles from the early 80s on the convergence of general threshold automata \cite{Dynamicsofpositiveautomatanetworks}.  Proving a lemma on the monotonicity of the threshold-based updating rule over the adoption vector\footnote{The threshold-based update is a deterministic mapping from the adoption vector at time $t$ to the adoption vector at time $t+1$ with input equal to the current adoption vector.}, and describing a key function in a more intuitive way,
we give a convergence bound that is better by a factor of 2 regardless of the network, and that is a arbitrarily-small fraction of the bound from \cite{Dynamicsofpositiveautomatanetworks} when the average degree is constant and the number of members of the network is large.\footnote{We note that these network properties are likely to hold for social networks.} Significantly, we give an upper bound on the convergence time that is a small linear multiple of the number of edges (or links) in the network (regardless of the degree distribution of nodes in the network).\footnote{
In both cases, some extensions to weighted variants (which may describe that some neighbors adoptions are more significant to an individual than others, or which may be convenient to model homophily-that people are more likely to be influenced by social contacts that are similar to them- are possible).}\\

\noindent \textbf{Connections to notions of seeding influence in Computer Science. } The spread and ``optimal seeding" of influence has been studied
extensively in the theoretical computer science literature for spread
mechanisms where a single infected contact is enough to cause conversion
(see, for example, the canonical work of Kempe, Kleinberg and Tardos
\cite{KempeKleinbergTardos} that has been cited over 1,400 times).
This model has attracted great interest in both academic and industry research communities for its connections to
viral marketing; as the body of observable data on how purchases
and product valuations flow through networks expands, companies
with products to sell and e-advertising space to allocate are
eager to learn as much as possible about how to market in this
new landscape. Analogously, as Axsen and Kurani suggest, we
seek to explore how green behaviors may be marketed:
what is fundamentally new about our work is addressing
the influence maximization question with a strong notion of local thresholds,
and with an emphasis on long term stability of adoption
(both considerations essential to an honest extension of the moral-motivation model). \footnote{We note that a few results on a related theme of reversible threshold processes exist in the mathematics literature, but the focus there is on characterizing extremely special classes of networks that exhibit certain properties (for example, that they can be completely converted by a set of a particular size given a particular uniform threshold, etc.) not on optimization of influence within a given network. For example, see \cite{Irreversiblekthresholdprocesses} or \cite{spenceadamsirreversible}. The discrete math literature has also considered a problem of when iterative local computation in networks (specifically, agreeing with the majority of one's neighbors) will converge to the majority opinion; the focus there is on characterizing a specialized set of networks in which convergence to a homogeneous state that reflects the majority of the initial adoption vector is guaranteed \cite{ListentoyourN}. Again, this work is strongly divorced from our planning perspective. }\\

\noindent \textbf{Implications for Cost/Benefit Analysis.} The moral-motivation model in \cite{Nyborg2006351}
offers a theoretical foundation compatible with
temporary subsidy policies that have already been implemented in practice,
and that appear to be succeeding in establishing stable
participation in personally-costly pro-environmental
behaviors.
A motivating example in \cite{Nyborg2006351} is a subsidy structure
adopted in Oslo, Norway where greener winter-traction
tires were subsidized until a fixed level of the population
was using the green option, then the subsidy was removed.
Prior to subsidy, studded tires were strongly preferred despite
the availability of green alternatives: this had resulted in
particulate air pollution from road-surface abrasion.
The
use of studded tires vs. greener traction tires is not
visually obvious, so effects of social-sanctioning seem likely to be small.
When the subsidy was removed green adoption
decreased only slightly, and then seemed to remain stable: the dramatic drop-off
classical arguments would predict based on the strong pre-subsidy preference for studded tires was not observed. Further, why would policymakers adopt a plan of this form?
The choice of such a subsidy structure seems to indicate that
policymakers may already believe that using temporary subsidies to reach high levels of green adoption
can have effects beyond the term of the subsidy.

That temporary subsidy may be able to enforce a permanent change in behavior
expands the range of policy settings in which subsidy is
a rational approach. Subsidy is generally rational
when the cost of providing the subsidy annually is outweighed by the
benefits that result from the resulting behavior change annually. But if temporary
subsidy can be used to cause a permanent shift in behavior, then the cost of imposing the
temporary subsidy for some number of years should now be weighed against
all benefits accrued during the term of the subsidy and
\textit{in the future} due to the behavior change that results
from the temporary subsidy (discounted appropriately).
In particular, in settings where moral-motivation is in play,
the size of investments in subsidy policies and public education
campaigns that are ``rational" may be much larger than when the
future-benefit of permanent behavior change is ignored.

Environmental economics is deeply concerned with appropriately
valuing investments in environmentally-sound practices: if the benefits of these investments are
under-valued (which may happen frequently in attempts to apply classical metrics for evaluating
investment returns created to describe much simpler products), or the estimates of their costs are inflated, societies risk making
highly-inefficient choices about allocation of resources.

This paper points out that if decision rules applied by
individuals rely on local information, then the level of investment required to reach a virtuous equilibrium may be much less than predicted by the non-network model. There
exist classes of networks where taking advantage
of the network structure can significantly reduce the number of
individuals that must be subsidized to reach the high-adoption equilibrium. Thus, the cost to reach the virtuous equilibrium could be much lower than
is predicted by the non-network model.
As environmental and behavioral economists and sociologists work to understand why
individuals adopt pro-environmental behaviors and the body of observable data about
social networks continues to explode,
leveraging network characteristics in considering the costs and
benefits of encouraging behavior change could critically impact cost-benefit
analyses used to evaluate important policies.\\


\noindent \textbf{Model Richness: The emergence of heterogeneous equilibria.} In addition to providing an encouraging message about the potential for moral-motivation to act as
an ally in stabilizing green equilibria, our network model
allows for long-term heterogeneous outcomes
in behavioral adoption that are intuitively satisfying in their realism.

Exploring similar issues, the statistical mechanics literature has considered spread mechanisms in random networks where individuals are influenced by trends in adoption among their neighbors; if a high-percentage of their neighbors adopt a behavior, they are very likely to adopt it, etc (e.g. \cite{socialstructureandopinionformation}, \cite{Diffusionincomplexsocial}). The widely-studied version in \cite{socialstructureandopinionformation} relies on a combination of a mathematically-convenient update rule that individuals periodically choose a random neighbor to emulate, and the assumption that these neighbors are always, in some sense, a representative sample of the society. The process defined gives a random walk which must eventually converge to an adoption pattern that is totally homogeneous: either the entire society participates or no one participates. This model behavior feels suspiciously simplistic, and insights based on such an approach feel brittle (for example, the finding that adding a single intransigent individual to a network- who will never alter their opinion- will unfailingly steer the entire society to their opinion \cite{socialstructureandopinionformation}). Concerned with this simplistic equilibrium behavior, recent work of Centola \cite{Centola2013} does manage to produce partial-adoption equilibria by assuming a secondary adoption threshold above which the sign of the benefits of adoption becomes negative. Such an assumption seems highly unnatural for describing norm spread.

Similarly, the traditional moral-motivation model from \cite{Nyborg2006351} can
result in partial-adoption equilibria (that is, a society-wide fraction of adoption that is strictly greater than 0 but strictly less than 1) only by
making some individuals more immune to moral-motivation than others: that is, non-adopters are evenly mixed throughout the population, and are precisely those individuals that don't respond to the pressure to conform to norms at modest rates of society-wide adoption.

Our network model, on the other hand, allows stable adoption patterns that are not homogeneous without relying on the moral callousness of all non-adopters: stable pockets of non-adoption may easily occur when certain portions of the network mainly talk to themselves and not to other portions of the network in which adoption is higher. This speaks to the flavor of the famous work of Nowak and May \cite{Nowak1992} on evolutionary games where introducing a network structure to repeated game play gives rise to persistent spatial heterogeneity. Anecdotal observations of the heterogeneous behavioral norms and locally-normalized values in our society also seem to recommend models of norm formation that are capable of producing a rich set of heterogeneous equilibria.\\

\noindent \textbf{Observability of Behavior and Diversity of Local Influences.} 
Egoistic moral-utility benefits associated with
conforming to locally-embraced green norms could also help to stabilize
observable pro-environmental behaviors (like curbside recycling, avoiding littering,
purchase of fuel-efficient vehicles, use of public transportation, etc). Further,
many anecdotes offered to illustrate social sanctioning have a strong local flavor (``your neighbor judges you for not recycling", etc).
If dis-utility experienced by non-adopters is non-decreasing in the fraction of local
adoption (a reasonable-seeming hypothesis) then incorporating this effect does not
change the form of the function we analyze in the following sections.

\section{The Moral-Motivation Model: Extending to Networks} \label{modeldef}
We first describe the basic moral-motivation model studied in \cite{Nyborg2006351}, then explain an equivalent threshold-based
decision rule, then extend this decision rule to a network.

In \cite{Nyborg2006351}, each individual $i$ has a
function $f_i(p)\geq 0$ of the fraction of society-wide adoption $p$ that describes the
moral utility benefit they receive for adopting the green behavior.  This is a non-decreasing function.\footnote{This assumption makes quantitative Frey and Stutzer's observation that, ``Individuals who believe that others will contribute to a public good tend to contribute more themselves." \cite{Environmentalmorale}, or see \cite{Socialcomparisonsandprosocial}.} See Figure 1 below.
The individual may also experience a small environmental-quality benefit $\beta$ as a result of
their green behavior choice (this is specified to be additive in \cite{Nyborg2006351}).
The sum of these benefits is weighed against the individual cost of adopting the green behavior $c$. This small environmental benefit $\beta$ from $i$ adopting is also accrued by all other members of the society.

\begin{figure}[ht!]\label{fig:moralutil}
  \centering
  \fbox{
  \begin{minipage}{12cm}
    \includegraphics[trim = 30mm 125mm 80mm 105mm, clip,
width=12cm]{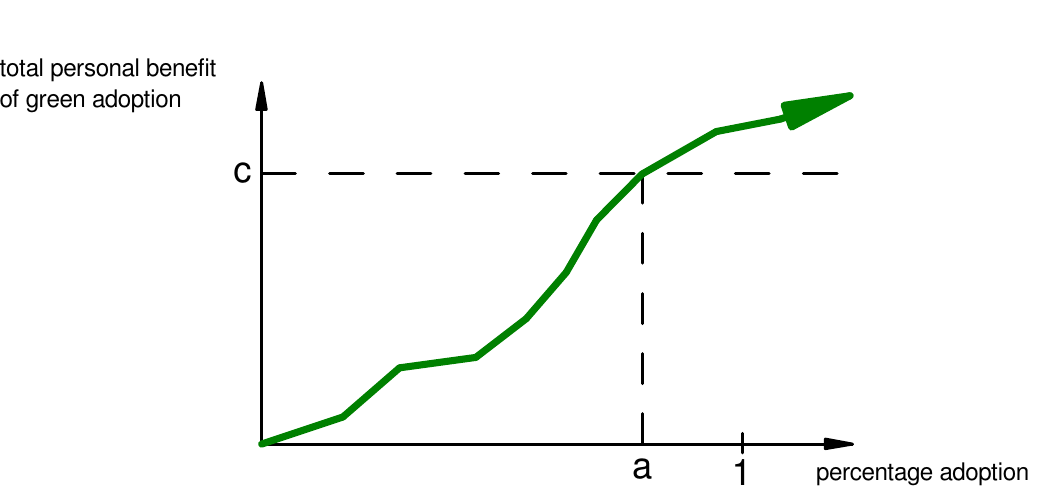} \caption{At the critical level of green-adoption \textit{a}, the sum of the moral utility benefit and the personal environmental benefit of switching to green behavior overcomes the personal cost, \textit{c}, associated with green behavior.}
\end{minipage}}
\end{figure}

\noindent The focus in \cite{Nyborg2006351} is to understand behaviors where both
\begin{itemize}
\item $f_i(0)+\beta<c$ : if no other members of society adopt the green behavior, then individual $i$ will not adopt the green behavior.
\item For at least some individuals $f_i(1)+\beta>c$ :  at a high rate of adoption at least some individuals will adopt even if they ar not directly subsidized to do so.
\end{itemize}
A temporary intervention (like a subsidy of value $s$ for adopting green behavior), can be introduced so that $f_i(0)+\beta+s>c$.  Thus, an individual will adopt the green behavior during the subsidy period.  At the time the subsidy is removed there is some society-wide adoption rate $p'$: if $f_i(p')+\beta>c$ then $i$ will continue to engage in the green behavior. Otherwise, $i$ will stop engaging in the green behavior. When the subsidy is discontinued, individuals will repeatedly perform updates based on this decision-rule: some erosion of adoption rate may occur.

In \cite{Nyborg2006351}, dynamics are studied for the case of uniform $f_i(\cdot)$ under the assumption that individuals update in a random order. These assumptions give a system that conforms to replicator dynamics: the convergence rate is proportional to the net benefit associated with adoption.  The replicator-dynamics convergence analysis in \cite{Nyborg2006351} relies critically on the assumption that, under any current level of adoption, the personal benefit of adoption is identical for every individual.\\

\noindent \textbf{Moral-Motivation in Networks.} We define an extended model that applies the local decision rule from \cite{Nyborg2006351} to each individual in a network.

\begin{itemize}
\item \textbf{Input:}\\
A network (or ``graph") consisting of nodes and edges, $G=(V,E)$. Each node in $V$ is in one of two states: \textit{Green} or \textit{Brown}. We associate the \textit{Green }state with value 1, and the \textit{Brown} state with value 0.

Each node $i$ has a critical susceptibility fraction $\alpha_i$.\footnote{ If $\alpha_i>1$ we say $i$ is intransigent (following the statistical mechanics literature that considers individuals of fixed opinions \cite{socialstructureandopinionformation}); $i$ will never adopt green behavior without subsidy.} This fraction is the minimal percentage of the neighbors of $i$ that must be \textit{Green} before $i$ will adopt the green behavior (without being subsidized), namely:
\[
\alpha_i = \text{argmin}_{p} \{p: f_i(p)+\beta > c\}.
\]
\textit{Notice that given $G$ and $\alpha_i$, we can immediately obtain a \textbf{cardinality threshold}, $b_i$, for adoption. Let $\delta(i)$ denote the set of $i$'s neighbors in G.
Then }$b_i=\lceil|\delta(i)|\alpha_i\rceil$
\textit{is the minimum number of green neighbors that will ensure $i$ adopts when no subsidy occurs.}

\item \textbf{Dynamics:} for each time step $t$, every node $i$ checks how many of his neighbors are \textit{Green}. If more than $b_i$ neighbors of $i$ are \textit{Green}, $i$ is set to \textit{Green} in time step $t+1$.  Otherwise, $i$ is set to \textit{Brown} in time step $t+1$. This update process iterates, giving a binary \textit{adoption vector} $x(t)\in \{0,1\}^{|V|}$ that evolves deterministically in time.\footnote{Our choice of concurrent updating, rather than the randomly-ordered updating explored in \cite{Nyborg2006351}, follows the extensive body of work on analyzing repeated game play in network settings that began in the early 90s in work such as \cite{Nowak1992}.}
\begin{figure}[ht!]
  \centering
  \hspace{8mm}\fbox{
  \begin{minipage}{15cm}
    \includegraphics[trim = 35mm 120mm 80mm 120mm, clip,
width=10cm]{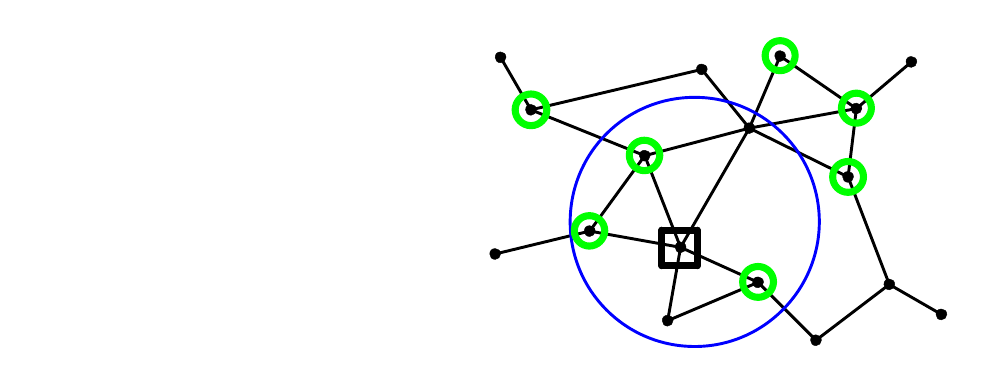} \caption{\textbf{Adoption in a local neighborhood determines behavior in the next time step.} Green adoption in the present time step is indicated by thick green circles. To decide adoption in the next time step, each node checks the percentage of adoption within his set of  neighbors. For example, the boxed node will adopt in the next time step if he has threshold $\alpha\leq 0.6$. }
\end{minipage}}
\end{figure}
\vspace{-4mm}

\item \textbf{Space of Actions:} Choose a set of nodes to temporarily subsidize (these nodes will be \textit{Green} regardless of their neighbors behavior until the subsidy is removed).

\item \textbf{Goal:} When the subsidy is removed, the long-term behavior of the adoption vector exhibits a high fraction of green behavior.
\end{itemize}


\noindent \textbf{In networks, stability of adoption depends on spatial distribution.} An important difference between the traditional moral-motivation model and our network version is in the ease of describing the long-term conversion achieved by subsidizing a particular set of nodes.  In the traditional model, if the individuals have uniform threshold ($\alpha_i$ is constant) the stability of conversion can
be determined exactly by whether or not the critical level of adoption has been exceeded during the term of the subsidy. In the network model describing the long-term stability of a subsidy strategy must also consider the spatial distribution of the adopting nodes.  A pattern of adoption and the configuration of network connections will together either maintain a high level of green behavior or allow green adoption to slowly erode. This added complexity makes description of the long-term effects of a temporary subsidy significantly more challenging. \\

\noindent\textbf{Natural notions of temporary subsidy in the network model:}

\begin{itemize}
\item \textbf{temporary (temp)}: The planner chooses a set of nodes to subsidize continuously until growth in green adoption stops, then the subsidy is removed.
\item \textbf{fixed-duration (fd)}: The planner chooses a set of nodes to subsidize for $d$ consecutive time steps, then the subsidy is removed.
\end{itemize}

\noindent When $d\geq |V|$ we will show that fixed-duration is a special case of temporary.
We list these models separately since we will prove much stronger results in the temporary case.
For each of these notions of temporary subsidy there are two natural planning problems to consider:\\

\noindent\textbf{Problem 1: Min-cost Complete Conversion (MCC)} What is the smallest number of nodes that can be subsidized to convert the entire network\footnote{Precisely, convert all nodes with $\alpha_i\leq 1$ permanently to green behavior; nodes with $\alpha_i>1$ never adopt without direct subsidy.} permanently to \textit{Green} behavior? We consider both tempMCC and fdMCC. \footnote{The assumption of uniform node-subsidy cost is adopted for ease of exposition, but our Integer Programs in Section \ref{sec:intprograms} can be trivially altered to accommodate non-uniform costs.}\\

\noindent \textbf{Problem 2: Budgeted Maximum Conversion (BMC)} Given budget of $k$ nodes to subsidize, what is the maximum percentage of the network that can be permanently converted to \textit{Green} behavior? We consider both tempBMC and fdBMC.\\
\vspace{-2mm}
\subsection{Summary of Results on Stability, Convergence, and Hardness}
For all 4 problem variants, we formulate time-indexed Integer Programs that compute the precise optimal set of seeds for stable green adoption.  The correctness of these optimal seed sets depends critically on several theorems we prove about convergence of the evolving adoption vectors (convergence from an arbitrary initial adoption state). For the case of temporary subsidy (temp) we show convergence to a single stable adoption vector. For fixed-duration (fd) subsidy we show convergence to a 2-cycle (in which two adoption vectors -that may be identical- alternate). Further, we prove upper bounds on the number of time steps the adoption vector (which evolves in the exponentially large space $\{0,1\}^{|V|}$) requires to converge.  In the worst case this upper bound is linear in the number of connections (edges) in the network (or equivalently, quadratic in the number of individuals in the network). These upper bounds determine the number of variables in our IP formulations.
Applying fast off-the-shelf solvers then allows practically-efficient measurement of optimal seed sets provided that the network (and its resulting IP) is not too large.

Our method of computing optimal sets is not efficient in a rigorous sense \footnote{According to the standard notion of efficiency in computer science: an algorithm is considered efficient if there exists a polynomial function $F(\cdot)$ with the property that the running time of the algorithm on any input $I$ is bounded above by the function $F(\cdot)$ applied to the size of $I$.} (currently, no rigorously-efficient method for solving general Integer Programs is known). In fact, we prove that no efficient methods to find optimal seed sets exist \footnote{Unless a ubiquitous assumption in the computer science literature is untrue, though technically the validity of the assumption is an open problem.}
by reductions from Set Cover and Maximum Coverage Models studied in the computer science theory literature. Further, a significant notion in computer science theory is that beyond proving that an optimal solution for a problem
cannot be found by any efficient algorithm, a class of problems may be proved \textit{hard to approximate}. This formalizes the idea that
the class of problems\textit{ cannot even be solved approximately by an efficient algorithm}. Consider a class of problems for which all feasible solutions for any input instance have non-negative value. The following is a widely-embraced definition in the literature (see \cite{Williamson:2011:DAA:1971947}):

\begin{definition}
An algorithm is a \textbf{$\beta$-approximation}
if, for every input instance, the algorithm produces in polynomial-time a feasible solution of
value at least $\beta$ times the optimal possible value given the input.
\end{definition}

For example, for a maximization problem, an efficient algorithm that always solves a problem perfectly is called a 1-approximation, an efficient algorithm that always returns a solution of value at least 1/2 the optimal value is a (1/2)-approximation, etc. For a minimization problem, an efficient algorithm that always solves a problem perfectly is called a 1-approximation, an efficient algorithm that always returns a solution of cost at most twice the optimal cost is called a (2)-approximation, etc.

The following table summarizes our results on the convergence states, upper bounds on maximum convergence times, and hardness of approximation for the four problem variants.\\

\noindent \textbf{Results on Stability, Convergence, and Hardness:}\\
\noindent \begin{tabular}{|l|l|l|}\hline
& \textbf{temporary (temp)} & \textbf{fixed-duration (fd)} \\ \hline
\textbf{Min-Cost} & converges to: stable adoption vector  & converges to: stable adoption vector   \\
\textbf{Complete}            & convergence time bound: $2|V|$        & convergence time bound: $d+2|E|+|V|$   \\
\textbf{Conversion}            & Hardness: $\Omega(ln (|V|) )$)       & Hardness: $\Omega(ln (|V|) )$          \\ \hline
\textbf{Budgeted}& converges to: stable adoption vector     & converges to: 2-cycle   \\
\textbf{Maximum}           & convergence time bound: $2|V|$ & convergence time bound: $d+2|E|+|V|$            \\
\textbf{Conversion}           & Hardness: $<1-\frac{1}{e}$ $\approx 0.632$    & Hardness: $<1-\frac{1}{e}$ $\approx 0.632$          \\ \hline
\end{tabular}

\vspace{3mm}

Notably, in the temporary subsidy model our upper bounds on convergence time of the adoption vector are linear in the number of individuals in the network (though the adoption vector evolves in an exponentially large space). In the more general fixed-duration subsidy model we give an upperbound on convergence time that is linear in the number of connections (edges) in the network. We note that the average degree in social networks is often bounded by a constant: for such networks all upper bounds given are linear in the number of individuals in the network. We also note that considering a network which is a simple line shows that the best possible upper bound that could exist on adoption vector convergence time is $|V|$: for social networks, at most a constant factor improvement could be given in the upper bounds we prove. The formal statement of these results and their proofs are included in the appendix.

For Min-cost Complete Conversion, we prove that no efficient algorithm is a $O(ln (|V|) )$-approximation. This is by reduction from the Set Cover problem (shown to be $\Omega(ln (|V|) )$-hard to approximate in the work of Feige \cite{Feige:1998:TLN:285055.285059}).  For Budgeted Maximum Conversion, we prove that no efficient algorithms achieves more than a $(1-\frac{1}{e})$ fraction of the optimal value. This is by reduction from the Maximum-Coverage Problem. The details of these reductions can be found in the appendix. These rather daunting hardness results for network Moral-Motivation point out that the well-studied ``Linear Threshold Model" of Kempe, Kleinberg, and Tardos (a tangent to the main results from \cite{KempeKleinbergTardos}) in the CS theory literature represents an very special
case of threshold-based transmission. Kempe, et al. employ a very specialized assumption on the form of uncertainty in threshold values to establish the property of submodularity; this property results in the success of certain approaches to solution construction.  Unfortunately, these approaches can fail dramatically for even slight variations on the form of uncertainty.

\section{Contrasts with non-network Moral-motivation }\label{sec:contrasts}
In \cite{Nyborg2006351} the authors describe a number of properties of the original moral motivation model. First note that our model generalizes the original moral-motivation model: the non-network model corresponds exactly to the special case of our model where $G$ is a complete graph.\footnote{The complete graph on a set of nodes $V$ has an edge between every pair of nodes in $V$. The complete graph on $|V|$ nodes is denoted $\textbf{K}_{|V|}$.}  How much can the properties of the model vary when the graph is not complete (that is, in our network moral-motivation model)? The answer is that the properties may be dramatically different.\\

  \fbox{
  \begin{minipage}{15cm}
  \vspace{2mm}
   \textbf{Original Moral-motivation (complete graph $G= \textbf{K}_{|V|}$):}

   \begin{enumerate}
   \item \textbf{The long-term adoption rate resulting from a subsidy depends only on the number of individuals subsidized.}

   \item \textbf{When $\alpha_i$ is uniform, achieving $100\%$-adoption requires subsidizing the critical-susceptibility fraction ($\alpha_i$) of the entire society.}\footnote{For the non-uniform threshold case: there exist examples where no subsidy of less than a $\max_i \{\alpha_i \}$ fraction of the society is sufficient for achieving $100\%$-adoption.}

   \item \textbf{When $\alpha_i$ is uniform, there are two stable equilibrium: no adoption and $100\%$-adoption. Heterogeneous stable equilibria can be obtained only when $\alpha_i$ is not uniform.}
   \end{enumerate}
   \textbf{Network Moral-motivation (general $G$):}
   \begin{enumerate}
 \item \textbf{The long-term adoption rate resulting from a subsidy depends strongly on its spatial placement in the network.}

 \item \textbf{There exist networks with uniform $\alpha_i$ in which $100\%$-adoption can be achieved by subsidizing a vanishingly-small percentage of the critical susceptibility fraction of society.}

 \item \textbf{Networks may have many stable heterogeneous equilibria in addition to 2 homogeneous equilibria, even when $\alpha_i$ is uniform}.

 \item \textbf{Even when a subsidy set $S$ can achieve $100\%$-adoption, almost every subsidy set of the size $|S|$ can fail to do so (sometimes miserably).}\\

   \end{enumerate}
\end{minipage}}

\vspace{2mm}
To make the above statements precise, we demonstrate classes of networks where the properties associated
with network moral-motivation depart strongly from the original version. We examine asymptotic classes of examples to illustrate the severity of the departure possible.  Later, in Section \ref{sec:computational}, these extreme departures
for pathological networks motivate our
exploration of the magnitude of departures for graphs that have been studied as proxies for social networks.\\

\noindent \fbox{
  \begin{minipage}{16cm}
\begin{theorem} \label{contrast1}(Moral-motivation in Networks)
   \textit{Two subsidy sets of the same size may have widely-different long-term adoption rates. This is true even when the $\alpha_i$ are uniform.}
\end{theorem}
\end{minipage}}

\vspace{3mm}

\noindent \fbox{
  \begin{minipage}{16cm}
\begin{theorem} \label{contrast2} (Moral-motivation in Networks)
  \textit{ For any $\epsilon>0$, there exists a network with uniform $\alpha_i$ in which stable conversion to the $100\%$-adoption equilibrium can be accomplished by subsidizing at most $\epsilon|V|\alpha_i$ individuals. }
\end{theorem}
\end{minipage}}

\vspace{3mm}

\noindent \fbox{
  \begin{minipage}{16cm}
\begin{theorem} \label{contrast3}  (Moral-motivation in Networks)
 Networks may have many stable heterogeneous equilibria, even when $\alpha_i$ is uniform.
\end{theorem}
\end{minipage}}

\vspace{3mm}

\noindent \fbox{
  \begin{minipage}{16cm}
\begin{theorem} \label{contrast4}  (Moral-motivation in Networks)
\begin{itemize}
\item Given any $\epsilon>0$ there exists a network which can be converted to $100\%$ adoption by a set $S$ but where a randomly-selected set of size $|S|$ fails to achieve $100\%$ adoption with probability at least $1-\epsilon$. This is true even when the $\alpha_i$ are required to be uniform.

 \item When the $\alpha_i$ are allowed to be non-uniform, given an additional parameter $\gamma>0$, there exists a network where some subsidy set $S$ gives $100\%$-adoption, but the expected adoption resulting from a random $|S|$-size set is $<\gamma$.
     \end{itemize}
\end{theorem}
\end{minipage}}

\vspace{3mm}

\noindent To prove these theorems we introduce several classes of networks.\\

\begin{figure}[ht!]\label{fig:trianglewithlegs}
  \centering
  \fbox{
  \begin{minipage}{12cm}
    \includegraphics[trim = 40mm 100mm 60mm 105mm, clip,
width=8cm]{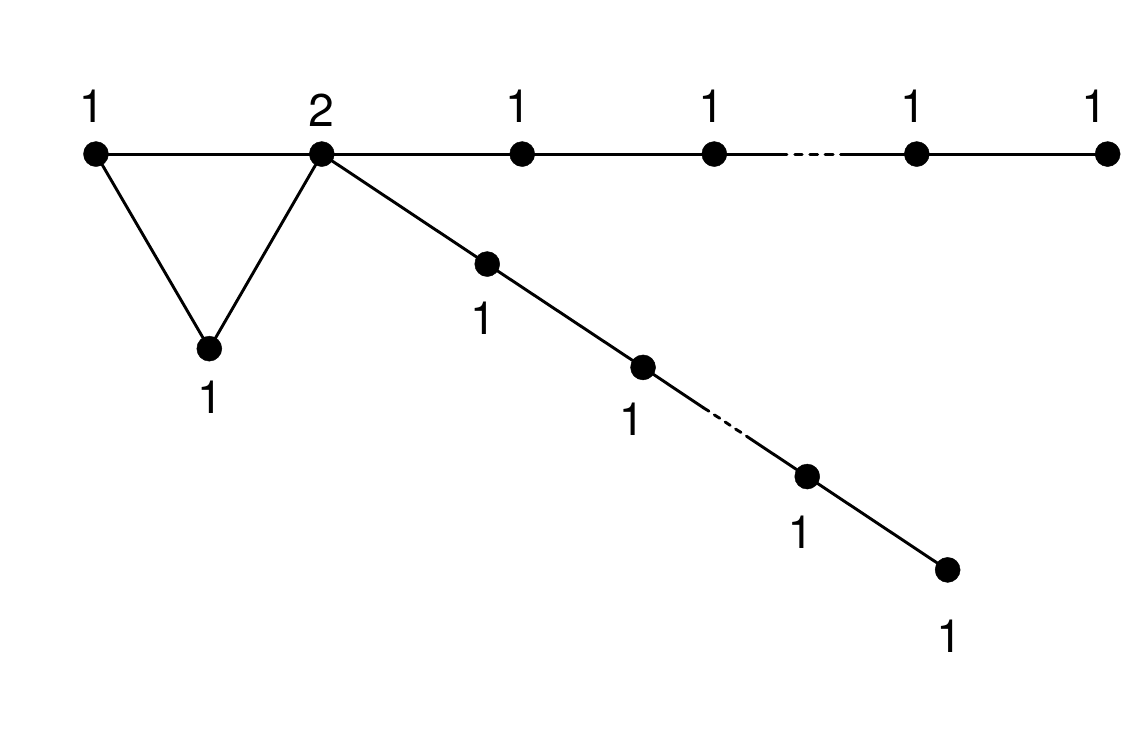} \caption{\textbf{\textit{Class 1:}} Resulting cardinality thresholds ($b_i$) are marked. Subsidizing any single node in the triangle for at least 3 time steps results in $100\%$-adoption, but subsidizing any single node not in the triangle (for any number of time steps) will result in long term adoption strictly less $50\%$.}
\end{minipage}}
\end{figure}

\noindent \textbf{\textit{Class 1:}} $G$ is composed of a triangle on three nodes, with two ``legs" each containing $n$ nodes (where the first such node in each leg is a common node from the triangle). Let $\alpha_i=1/2$ for all $i\in V$. See Figure \ref{fig:trianglewithlegs}.

\begin{itemize}
\item \textit{Proof of Theorem \ref{contrast1}:} As explained in Figure \ref{fig:trianglewithlegs}, two subsidy sets of size 1 have widely-different long-term adoption rates for Class 1 (which has uniform $\alpha$).

\item \textit{Proof of Theorem \ref{contrast2}:} Regardless of the value of $n$, a network in Class 1 can be converted to $100\%$ adoption by subsidizing any single node in the triangle for at least three time steps. A network of Class 1 has $2n+1$ nodes. Thus, for any $\epsilon>0$, we can choose $n'$ sufficiently large so that $1 <  \epsilon (2n'+1)\frac{1}{2}=\epsilon|V|\alpha$. We obtain an infinite class of examples (for $n\geq n'$) which satisfy Theorem \ref{contrast2}.

\item \textit{Proof of Theorem \ref{contrast4} (first half):} For a network in Class 1 there are three possible subsidy sets of size 1 which result in $100\%$ adoption if subsidy is maintained for three timesteps. There are $2n-2$ sets of size 1 that result in strictly less than $50\%$ adoption (regardless of how long the subsidy is maintained).  Thus, while $100\%$ adoption is achievable by subsidizing a single node, for any $\epsilon>0$, there exists sufficiently large $n$ so that the probability of randomly selecting a single node that achieves $100\%$ adoption is $3/(2n+1)<\epsilon$.
\end{itemize}

\noindent \textbf{\textit{Class 2:}} $G$ is composed of a triangle on three nodes, with $n$ ``legs" of length 3 attached to a common node of the triangle (the first node of each such leg is the common node in the triangle). All nodes of degree 1 or 2 have $\alpha_i=1/2$. The node of degree $n+2$ has $\alpha=2/(n+2)$. See Figure \ref{fig:trianglewithbristles} below.

\begin{figure}[ht!]\label{fig:trianglewithbristles}
  \centering
  \fbox{
  \begin{minipage}{12cm}
    \includegraphics[trim = -65mm 23mm -20mm 15mm, clip,
width=7cm]{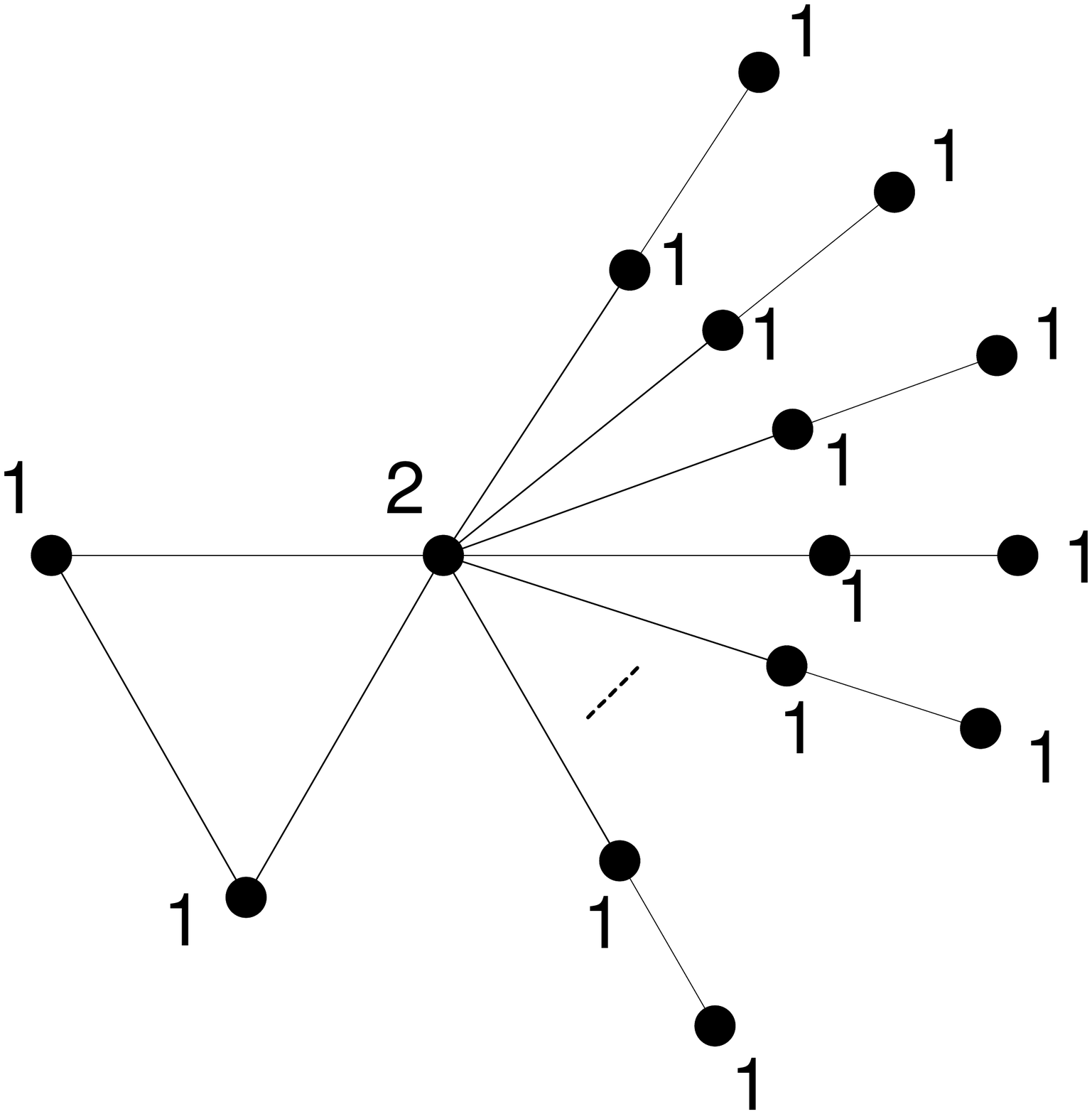} \caption{\textbf{\textit{Class 2:}} Resulting cardinality thresholds ($b_i$) are marked. Subsidizing any single node in the triangle for at least 3 time steps results in $100\%$-adoption, but subsidizing any single node not in the triangle (for any number of time steps) will result in long term adoption rate of at most $2/(2n+3)$.}
\end{minipage}}
\end{figure}

\begin{itemize}
\item \textit{Proof of Theorem \ref{contrast3}:} For networks of Class 2 with $n\geq 1$, there are both homogeneous and heterogeneous stable equilibria.  $100\%$ adoption is a stable equilibrium, $0\%$ adoption is stable equilibrium, and there are $n$ stable heterogeneous equilibria of the form: the two outer nodes in a leg are adopters and no other nodes adopt.

\item \textit{Proof of Theorem \ref{contrast4} (second half):} For a network in Class 2 there are three possible subsidy sets of size 1 which result in $100\%$ adoption if subsidy is maintained for three timesteps. There are $2n$ sets of size 1 that result in $(2/(2n+3))$ adoption (regardless of how long the subsidy is maintained).  Computing the expected adoption rate achieved by selecting a random set of size 1:
    \[
    \text{Expectation }=\frac{3}{2n+3}(1)+ \frac{2n}{2n+3}\Big(\frac{2}{2n+3}\Big)\leq \frac{10n+9}{(2n+3)^2}
    \]
    Thus, while $100\%$ adoption is achievable by subsidizing a single node, for any $\gamma>0$, there exists sufficiently large $n$ so that the expected adoption rate achieved by selecting a random set of size 1 is strictly less than $\gamma$.

\end{itemize}

\section{Formulating Integer Programs}\label{sec:intprograms}

In this section we apply our results on convergence states and times to write down Integer Programs that perfectly calculate the optimal set
of individuals (nodes) to subsidize. Let $\delta(i)$ denote the set of neighbors of $i$ in $G$.  Recall the problem variants from Section \ref{modeldef}.

\subsection{temporary Min-cost Complete Conversion (tempMCC)}
Decision Variables:
\begin{itemize}
\item \textit{(Subsidy Variables)} for $i\in V$:
\[
 y_i = \begin{cases}
        1  & \text{ if node $i$ is subsidized from time step 0 to $|V|$}\\
        0  & \text{ otherwise}
        \end{cases}
\]

\item \textit{(Adoption Variables)} for $i\in V, t\in \{0,1,2,..., 2|V|\}$:
\[
 x_{it} = \begin{cases}
        1  & \text{ if node $i$ adopts at time $t$}\\
        0  & \text{ otherwise}
        \end{cases}
\]

\item \textit{(Subsidy Size)} a dummy variable $q$ describes the size of the subsidized set.\\
\end{itemize}
We wish to minimize $q$ subject to the following constraints.  First, the number of nodes that are chosen for subsidy is at most $q$:
\[
\sum_{i\in V} y_i \leq q
\]
During the period of the subsidy, a node $i$ is allowed to be an adopter only if either it is subsidized, or if at least $b_i$ of its neighbors are adopters during the previous time step. Due to the integrality of the \textit{Adoption} and \textit{Subsidy} variables this condition is enforced precisely by the following inequality:
\[
 x_{it}  \leq y_i+\frac{1}{b_i}\sum_{j\in \delta(i)} x_{j,t-1} \text{\hspace{8mm}for $ t \in \{0,1,2,..., |V|\},  i \in V$.}
\]

From the proof of Lemma \ref{growthphase}  in the appendix, assuming that the subsidy is maintained for $|V|$ time steps and is then removed gives the same behavior as if subsidy is removed immediately after the first step in which no 0 to 1 conversions occur (that is, assuming subsidy is maintained for $|V|$ time steps perfectly emulates the conditions we defined for temporary subsidy). After the period of subsidy ends, a node $i$ is allowed to be an adopter only if at least $b_i$ of its neighbors are adopters during the previous time step:

\[
x_{it}\leq \frac{1}{b_i}\sum_{j\in \delta(i)} x_{j,t-1} \text{\hspace{8mm} for $t \in \{ |V|+1,|V|+2, ..., 2|V|\}, i \in V$.}
\]

From Lemma \ref{erosionphase} in the appendix, imposing the above condition for $|V|$ time steps after subsidy is removed ensures that the process will have converged to a stable adoption vector. Forcing the adoption variables for the final time step to 1 ensures we only consider solutions which result in $100\%$ adoption. This is enforced by requiring that for all $i\in V$:
\[
x_{i,2|V|} \geq 1
\]

Combining these constraints we obtain the following IP that computes the smallest subsidy set which permanently converts the entire network.

\begin{align*}
\text{minimize } q \hspace{5mm} & \\
 \text{subject to} \hspace{5mm} &\\
  \sum_{i\in V} y_i &\leq q\\
   x_{it}           &\leq y_i+\frac{1}{b_i}\sum_{j\in \delta(i)} x_{j,t-1}\hspace{5mm}        \text{ for } t \in \{0,1,2,..., |V|\},  i \in V.\\
   x_{it}           &\leq \frac{1}{b_i}\sum_{j\in \delta(i)} x_{j,t-1}\hspace{5mm}                 \text{ for } t \in \{ |V|+1,|V|+2, ..., 2|V|\}, i \in V.\\
   x_{i,2|V|}& \geq 1 \hspace{5mm}                 \text{ for }  i \in V.
\end{align*}

\noindent\textbf{\textit{A technical note: }}a sequence of adoption variables that constitute a feasible solution for the IP may not perfectly match the evolving adoption vector of the process. In particular, a set of adoption variables may ``lag" the true adoption vector: the process effectively turns on (from 0 to 1) a node as soon as possible, while we only require that the IP not turn on (from 0 to 1) a node if it is not justified in doing so (the second and third sets of constraints ensure this). This discrepancy might seem disconcerting for a moment; we might worry that the IP could gain some advantage by lagging the process, and thus produce a subsidy solution that is apparently the best but in fact performs badly with respect to the true adoption-evolution process. The answer to this concern is Theorem \ref{monoton} from the appendix on the monotonicity of the process: when our goal is enlarging the set of adopters, there is never an advantage in delaying turning a node from 0 to 1.

The formulation of the IPs for the other 3 problem variants is highly similar (though the form of the constraint sets for the fixed-duration cases depend on the case-specific convergence bounds we prove in the appendix).  The three other model formulations (tempBMC, fdMCC, fdBMC) are included in the appendix.

\subsection{Intransigent Individuals, Weighted Neighbors, Differential Costs of Subsidy}
\textbf{\textit{Intransigence:}} The model can describe \textit{intransigent individuals} by letting the cardinality threshold, $b_i$, exceed the degree of the node, $|\delta(i)|$.  This threatens to make our Min-cost Complete Conversion Problems infeasible: intransigent individuals will clearly violate the constraint that they be adopters in the final time step. By doing a simple iterative search in the network, it is easy (and efficient) to identify a set of nodes that are explicitly intransigent or implicitly intransigent (they are intransigent once all explicitly intransigent nodes are removed from the network, or once other implicitly intransigent nodes are removied, etc).  This gives a \textit{network of intransigence}: the planner can never hope to influence these nodes to adopt without direct indefinite subsidy.  Removing the final time step constraint for nodes in this \textit{network of intransigence} gives a variation of Min-budget Complete Conversion in which we can compute the minimum-size subsidy set that permanently converts all nodes which are not in the \textit{network of intransigence}.\\

\noindent \textbf{\textit{Weighted Neighbors:}} Suppose that the adoption of some set of $i$s neighbors is more important to $i$ than the adoption of other sets of $i$s neighbors.  We might model this by weighting each neighboring adoption status (0 or 1) by a scalar before comparing to a threshold $b_i$.  In particular, when symmetric weights are used to describe homophily of individuals (how alike a pair of individuals are) and weights are chosen from the set of integers $\{1,2,...,k\}$ for some constant $k$, nearly all our results go through immediately (with somewhat diluted, but still linear, upperbounds on convergence time for the fixed duration case). We mention this in reference to recent work in the mathematical sociology literature on how homophily appears to  contribute substantially to increased rates of healthy-behavior transmission in social networks. \cite{homophiliyinadoptionofhealth}\\

\noindent \textbf{\textit{Differential Costs of Subsidy:}} None of our convergence or monotonicity results require the cost of subsidizing nodes to be uniform. Trivial alteration of the Integer Programs specified allows perfect computation in the non-uniform case.

\section{Computational Results}\label{sec:computational}

The integer programs introduced can be solved for modestly-sized networks using off-the-shelf integer-programming solvers (we use CPLEX to solve models formulated in the AMPL language).  A large space of questions might be explored using these tools.  We defer more extensive computational exploration
to a later study, and concentrate here on preliminary computational observations related to the asymptotic results in Section \ref{sec:contrasts}.

 \begin{figure}[ht!]\label{fig:clusterrand}
  \centering
  \fbox{
  \begin{minipage}{12cm}
    \includegraphics[trim = 52mm 105mm 10mm 108mm, clip,
width=12cm]{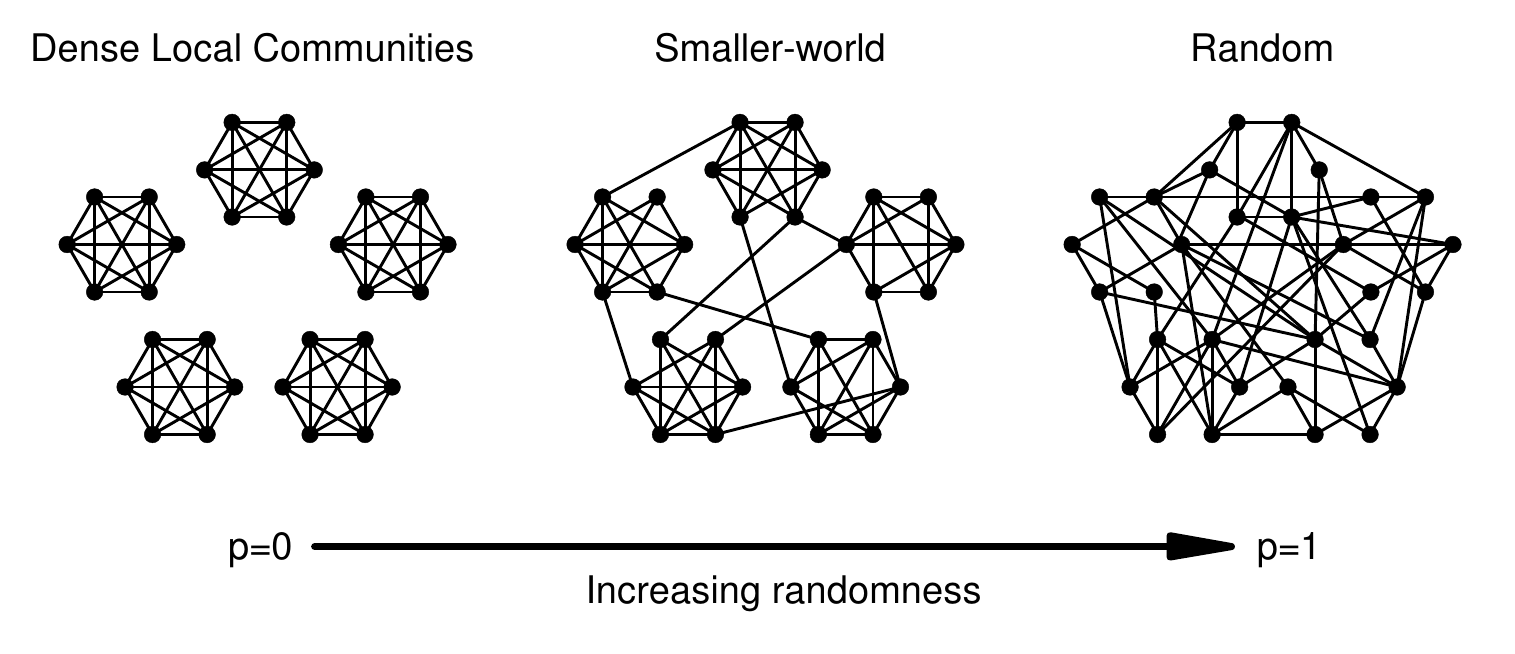} \caption{\textbf{Randomly-rewired clusters.} A random-rewiring scheme analogous to that in \cite{Watts1998} interpolates between dense local communities (small complete graphs) and a random network. Above, we start with 5 complete graphs each containing 6 nodes. For each edge in the graph we ``rewire" it with probability \textsf{p}: the edge is removed from the graph and replaced by a randomly-chosen edge between one of its endpoints and a random node from any community. We do this procedure for every edge in the leftmost figure, eliminating rare duplicate edges when they appear. When \textsf{p=0} no edges are rewired: this gives the leftmost figure. As \textsf{p} increases, some local ties are replaced by longer random ties, reducing the average path-length between pairs, but maintaining fairly high clustering coefficient. When \textsf{p=1} the procedure will give a random graph with low clustering coefficient, very low average path length (and closely-preserved average degree).}
\end{minipage}}
\end{figure}

\noindent \textbf{Networks that look like social networks.} Social Networks are generally characterized by high levels of local density (or high ``clustering coefficient"\footnote{Following \cite{Watts1998}, the clustering coefficient is defined as follows.  For each node $v$ in the graph, look at its set of immediate neighbors (not including itself). Let $k_v$ denote the size of this set. The largest number of edges that could exist between two members of this set is $k_v(k_v-1)/2$. Let $C_v$ denote the fraction of these possible edges that actually exist. The clustering coefficient of the graph is the average of $C_v$ over all nodes $v$.}) in combination with low average shortest-path length between randomly-selected pairs \cite{Watts1998}.  The canonical work of Watts and Strogatz \cite{Watts1998} demonstrated that starting from uniform lattices (rings and square lattices), random ``rewiring" of even a small percentage of the edges quickly gives rise to networks with these properties (qualitatively, even a very small percentage of random long ties results in a small-world). Similar rewiring schemes have been used widely in the sociology literature to create test networks to examine the spread of complex contagion (see \cite{Cascadedynamicsofcomplexcontagion} and \cite{Complexcontagionsandtheweaknessoflongties}).

In this section we describe computational observations for classes of networks with these properties that are composed of dense local communities, which are linked via a random rewiring procedure. Then we computationally investigate analogs of Theorem \ref{contrast2} and Theorem \ref{contrast4} for these networks.

\subsection{What percentage of subsidy can convert the whole network?}
\begin{figure}[ht!]
  \centering
  \fbox{
  \begin{minipage}{12cm}
    \includegraphics[trim = 48mm 80mm 65mm 90mm, clip,
width=12cm]{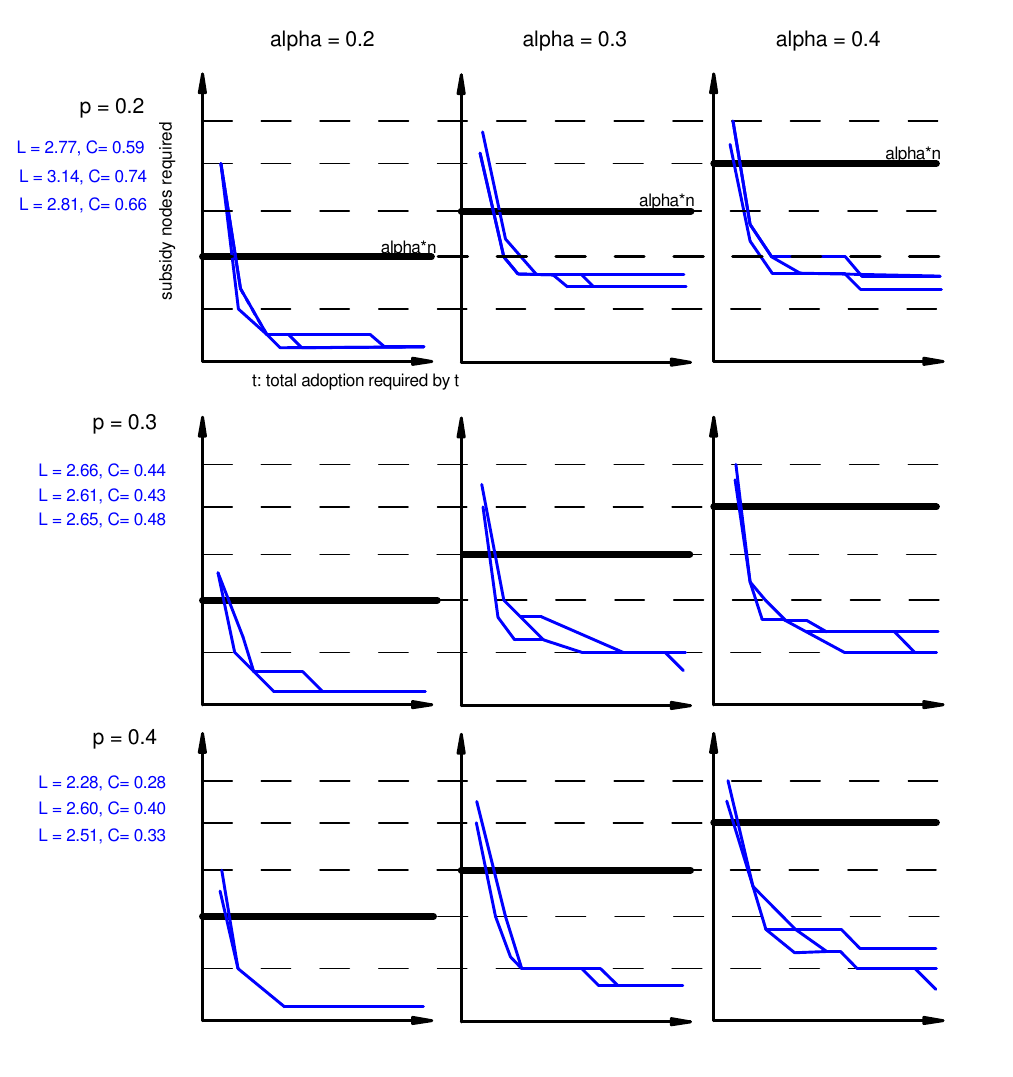} \caption{\textbf{Smaller subsidies suffice to reach $100\%$ adoption}: Unless we require that total adoption be reached very quickly, targeted subsidy of at most $\frac{1}{2}\alpha n$ nodes appears to suffice in this parameter range. Each row of the figure shows three runs for a fixed rewiring probability \textsf{p} in the 30-node graph described in Figure 5 (with the clustering coefficient \textsf{C} and average path length \textsf{L} listed for each run at left) for varying values of the threshold level $\alpha$.}
\end{minipage}}
\end{figure}
In Theorem \ref{contrast2} we saw that for some highly-pathological classes of networks (with uniform threshold $\alpha=1/2$), subsidizing a vanishing fraction of the critical percentage of the network (a vanishing fraction of $\alpha n$) would be sufficient to convert the entire network to adoption.  \\

\noindent \textit{\textbf{Question: } Is there a very small subsidy set that converts the entire network for networks that look like real social networks?}\\

Our observations for the rewired-cluster graphs of size 30 (as described in Figure 5) indicate that for a range of \textsf{p} values, giving rise to a range of (clustering coefficient, average path length) pairs, and a range of $\alpha$ values, the optimal size of a set that converts the entire network is less than $\frac{1}{2}\alpha n$. See Figure 6 for a graphical summary over a range of ($\textsf{p}, \alpha$)-pairs.

When we require that complete adoption in the network is obtained quickly the size of subsidy required increases, but only becomes larger than $\alpha n$ when we insist that complete adoption be reached after only one time step of subsidy.  Qualitatively, if local-norm spreading is given time to operate, the percentage of subsidy required to reach $100\%$ adoption decreases substantially from what is predicted by the non-network Moral-motivation model of \cite{Nyborg2006351}.


\subsection{How much does network-position-based targeting help?}

If we had no specific knowledge about the social network, we might optimistically hope that a randomly-placed subsidy in the network would perform reasonably well compared to the best-targeted subsidy of the same size. Qualitatively, we are hoping that the information deficit (lack of knowledge of the network) is not too costly in terms of the long-term adoption we are able to establish with our subsidy campaign.

\begin{figure}[ht!]
  \centering
  \fbox{
  \begin{minipage}{12cm}
    \includegraphics[trim = 55mm 80mm 40mm 80mm, clip,
width=12cm]{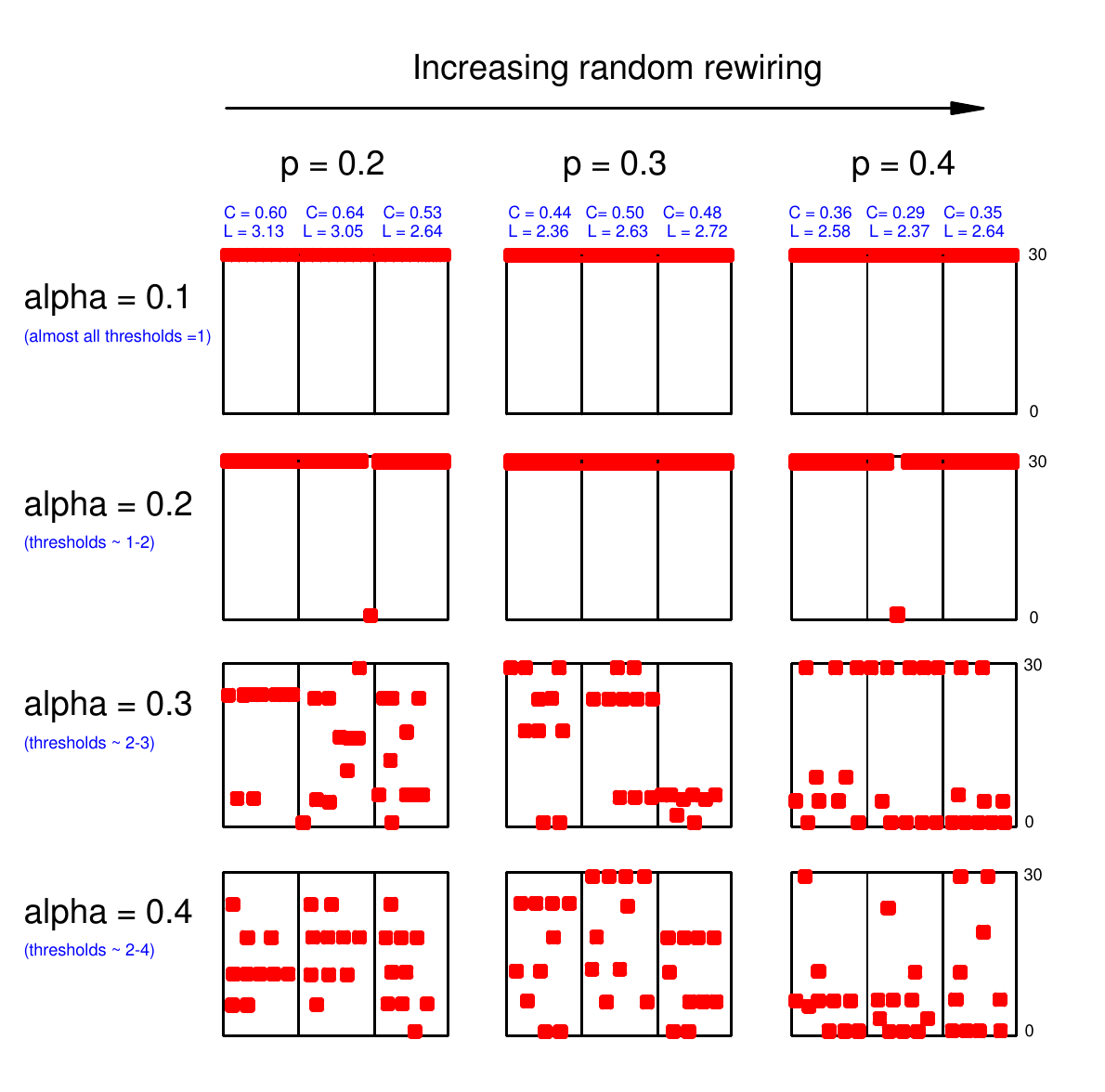} \caption{\textbf{As thresholds increase, the advantage of targeting increases}: When thresholds are small, randomly-selected sets of the same size as the optimal seeding set result in $100 \%$ long-term adoption. As thresholds increase, the average adoption resulting from a randomly-selected set decreases (though notably, this quantity has high variance, and some randomly-selected sets do result in $100\%$-adoption). This observation appears to hold for a range of rewiring probabilities, \textsf{p}: for each \textsf{p} the figure shows three randomly-rewired clusters of size 30.  For each randomly-rewired cluster, over a range thresholds, the performance of 10 randomly-selected seed sets (of size identical to the optimal seed set for total adoption) is observed. }
\end{minipage}}
\end{figure}
In Theorem \ref{contrast4} we saw that for certain classes of specialized networks, even when a small set exists that converts the entire network, a randomly-selected set of the same size could result in very low expected long-term adoption.  Qualitatively, this means that in certain classes of networks, careful targeting gives substantial gains in long-term adoption.\\


\noindent \textit{\textbf{Question:} For networks that look like real social networks, if $k$ targeted subsidies are sufficient to convert the entire network, how much adoption results from a randomly-selected $k$-set?}\\

In Figure 7 we observe that when thresholds are very small, randomly-selected sets perform similarly to the optimal seeding set. However, as thresholds increase (corresponding to when the percentage of local-adoption must be higher in order to surpass the personal cost of the green-behavior) we observe that randomly-located subsidies are substantially less effective on average. That is, the importance of targeting appears to increase as thresholds increase.  We note that some randomly-selected sets do perform very well in terms of long-term adoption, though we suspect that how often this happens may be influenced by the number of initial clusters (in this case 5) and the size of the optimal seeding sets (which are relatively small in the examples we observe). For the parameter range we examine though, small sets that result in high long-term adoption do not appear to be highly-rare. We note that the heuristic strategy of randomly selecting many sets of small size and choosing a good one does require knowledge of the social network.

Notice in Figure 7 that the performance of the randomly-selected sets often corresponds to multiples of 6: this is due to green-behavior stabilizing in several local communities but not being able to spread through the rewired ties into the remaining communities.


\textit{Comment:} We consistently observed that for a fixed network, solve times were much slower for large values of $\alpha$ (where ``large" is above $1/2$).

\vspace{-2mm}

\section{Conclusion}
To explore how the spread of pro-environmental behaviors can be best encouraged in social networks,
we extend the Moral-motivation model from \cite{Nyborg2006351} to networks. The green behaviors we are interested in spread in a very different way from disease or information, and motivate a complex optimization problem.
To examine natural planning questions about seeding the spread of green-norms, we create a practical computational tool that relies on new results we prove on stability and convergence of an evolving spatially-heterogeneous vector of adoption. For specialized classes of networks, targeting subsidy based on the social network and adoption thresholds yields huge gains over strategies that specify only the size of the subsidized set.

Our computational experiments in networks which replicate the high clustering and small-world path lengths characteristic of real social networks indicate that milder, but still substantial, advantages may be gained by careful subsidy targeting. Targeting subsidies based on knowledge of social-network structure could result in more long-term green adoption, or in substantial reduction in the costs of stabilizing wide-spread participation in pro-environmental behaviors.  The subsidy size required to establish stable society-wide adoption is predicted to be much smaller than under the non-network model (at most half the size for the modest examples we explore).
Further, in networks that resemble real social networks, the relative advantage of social-network-based targeting appears to increase as thresholds increase.
For green behaviors where moral motivation applies, good information about social-network structure and targeted subsidy based on this information may increase the benefits and reduce the costs of
subsidy and public education campaigns that aim to establish long-term participation in environmentally-friendly behaviors. A rigorous understanding of the local-norm-driven nature of marketing green adoption could critically impact the future cost/benefit calculations that inform public investment aimed at achieving sustainable practices.

We have just scratched the surface of the computational exploration that is possible with the IP tool we have introduced here.  A study on the properties of sets of high-influence with respect to the threshold-based spread mechanism described here is underway; we expect significant departures from the qualitative properties of sets of high influence under information-spread mechanisms. There are also a number of natural questions on the robustness of seeding strategies to noisy specification of the network, and how stability of high-green-adoption equilibria is threatened by evolving network structure, increasing personal costs of green behaviors, etc.

\newpage

\bibliographystyle{}
\bibliography{masterbib}

\section{Appendix: Theoretical Results on Monotonicity and Convergence}
In this section we establish several theorems about the Moral-motivation model in networks that will be required in our Integer-Program formulations. A prime issue in the formulation is the convergence time and convergence states of an adoption vector: how many time steps of the update process are necessary in order to
describe the long-term adoption rate?

For adoption vector $x(t)\in \{0,1\}^{|V|}$, let $A_{x(t)}$ denote the set of indices $i$ for which $x_i(t)=1$.\\

\noindent \fbox{
\begin{minipage}{16cm}
\begin{theorem} \textit{(Monotonicity of the future adoption set in the current adoption set)}\label{monoton}
Let $x$ and $x'$ denote two different adoption vectors that are evolving in time under identical subsidy conditions. For arbitrary $t$, if $A_{x(t)}\subseteq A_{x'(t)}$, then for all $T > t$, $A_{x(T)}\subseteq A_{x'(T)}$.
\end{theorem}
\end{minipage}}\\

\noindent\textit{Proof:} We prove the theorem by induction, starting at time step $t$.  From the assumptions of the theorem, $A_{x(t)}\subseteq A_{x'(t)}$.  Induction hypothesis: suppose that for time step $T-1$ we have $A_{x(T-1)}\subseteq A_{x'(T-1)}$. Consider time step $T$. For each node $i\in V$:
if $i \in A_{x(T)}$, then it must be the case that at least $b_i$ neighbors of $i$ are in $A_{x(T-1)}$. Since all nodes which are in $A_{x(T-1)}$ are also in $A_{x'(T-1)}$ (by the induction hypothesis), at least $b_i$ neighbors of $i$ are in $A_{x'(T-1)}$. Thus, by the update rule for $i$, we get that $x_i'(T)=1$, so that $i\in A_{x'(T)}$.  $\Box.$

\subsection{Temporary Subsidy and Fixed Duration with $d\geq |V|$: $2|V|$ time steps are sufficient to converge to a stable adoption vector}
The analysis in this section is relatively simple because the growth of adoption (updates from 0 to 1) and erosion of adoption (updates from 1 to 0) are divided cleanly into separate phases. The growth phase occurs while the subsidy is maintained, and the erosion phase begins in the first time step in which subsidy is removed. We explain below.

We start with a trivial but useful observation. Suppose that the subsidy is either maintained indefinitely, or has already been removed. If there is a series of 2 time steps in which the adoption vector is unchanged, then it will never change in the future.  That is, if $x(t)=x(t+1)$, then for all $T\geq t$, we have $x(t)=x(T)$.\\

\noindent \fbox{
  \begin{minipage}{16cm}
\begin{lemma} \label{growthphase}(Growth Phase)
For any set of nodes $S\in V$: if $S$ is subsidized from time 0 until the first time-step $g$ in which no new adoptions occur, then $g$ is at most $|V|$. Further, if subsidy of $S$ is maintained indefinitely after $g$ (and no additional nodes are subsidized), no new adoptions occur after $g$.
\end{lemma}
\end{minipage}}\\

\noindent \textit{Proof:} While subsidy of $S$ is maintained there is no back-migration from 1 to 0 (recursively, all nodes that convinced any node to adopt are still adopters themselves while the subsidy is maintained). If there is a single step $g$ with no new node adoptions, growth of adoption must be over since the next new adoption that occurs, say in time step $q$, will be based on adoption vector $x(g-1)$ just as $x(g)$ was: since $x(g)$ has no new adopters, $x(q)$ will have no new adopters (contradicting the choice of time step $q$ as the next in which a new adoption happens). Thus, since at least one node newly adopts in each time step (starting from time 0), and no back-migration occurs while subsidy of $S$ is maintained, after at most $|V|$ timesteps adoption has stopped increasing. $\Box$.\\

\noindent \fbox{
\begin{minipage}{16cm}
\begin{lemma} \label{erosionphase}(Erosion Phase)
Let $x(r)\in \{0,1\}^{|V|}$ denote the adoption vector at the time step $r$ when subsidy is removed. If $x(r)=x(r-1)$, then
all future updates will be from 1 to 0, and the adoption vector will converge to a stable adoption vector within $|V|$ time steps.
\end{lemma}
\end{minipage}}\\

\noindent \textit{Proof:} Since $x(r)=x(r-1)$, no future updates from 0 to 1 will occur (as argued in the proof of the previous lemma). In particular, let $S'$ denote the set of all non-adopters at time $r$: all nodes in $S'$ will have value 0 in all future time steps. Setting $b'_i=\max\{\delta(i)-b_i,0\}$: view the process of erosion as a subsidy at nodes in $S'$ (to encourage converting to 0) which is maintained indefinitely and spreading via the network moral-motivation model with thresholds $b'_i$. The evolving adoption vector that results is exactly identical to the evolution of the adoption vector for the original process.
Applying the previous Lemma gives that a time step with no new 1 to 0 updates will occur within $|V|$ time-steps, and that after that time step the adoption vector will never change. $\Box$.\\

\noindent As a direct consequence of these two lemmas, we obtain:\\

\noindent \fbox{
\begin{minipage}{16cm}
\begin{theorem} (Convergence for tempMCC, tempBMC, and fd with $d\geq |V|$)
Consider a subsidy that is maintained for $|V|$ time steps (or until growth in new adoptions stops), and is then removed.  A total of
$2|V|$ time steps is sufficient for the adoption vector to converge to a stable equilibrium.
\end{theorem}
\end{minipage}}\\

\subsection{General fixed-duration case: $d+2|E|+|V|$ steps are sufficient to converge to a cycle of length at most 2}

After the subsidy is removed at time $d$, allowing how many more time steps to elapse will ensure that the adoption vector accurately gauges long-term adoption? Notably, unlike in the previous case, the growth and erosion of adoption may now proceed simultaneously in the network. In the previous case, we naturally got convergence to a stable adoption vector. When subsidy is fixed-duration, this is not guaranteed.  Consider the example given in the Figure below.

\begin{figure}[ht!]\label{fig:alternating}
  \centering
  \fbox{
  \begin{minipage}{16cm}
    \includegraphics[trim = -70mm 95mm 20mm 95mm, clip,
width=11.5cm]{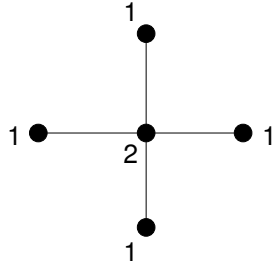} \caption{\textbf{(An alternating cycle of length 2 with average adoption $50\%$)} Here $\alpha$ is uniformly 1/2. Suppose that the planner can subsidize one node for one time step. Subsidizing a node of degree 1 gives long term adoption 0. Subsidizing the central node in time step 1: in time step 2, the central node looks at its neighbors status from time step 1 and thus stops adopting, while each degree-1 node sees that the central node adopted in time step 1 and so adopts in time step 2. In time step 3: each degree-1 node sees that the central node did not adopt in time step 2 and thus decides not to adopt in time step 3, while the central node sees at least 2 neighbors adopted in time step 2 and so adopts in time step 3. Thus, adoption in time step 3 is identical to adoption in time step 1. Since updates are deterministic, the adoption vector is in a cycle of length 2: every odd time step has only the central node adopting, every even time step has all nodes except the central node adopting. Thus, the optimal average adoption is $50\%$.}
\end{minipage}}
\end{figure}

It is not immediately clear that a meaningful concept of long-term behavior exists if the adoption vector does not converge. In the example in Figure \ref{fig:alternating}, an average rate can be computed because we notice that adoption has entered a small cycle (of length 2), but suppose that this does not happen.  Perhaps the adoption vector enters an exponentially-long cycle, or no cycle at all. This would endanger our scheme to optimize the adoption rate with a small Integer Program: the objective function might have exponentially-many terms. Fortunately, the example in Figure \ref{fig:alternating} is in essence as bad as possible: we can always capture the longterm average adoption as an average over two consecutive adoption vectors (provided we wait a sufficient number of time steps before computing this quantity).\\

\noindent \fbox{
\begin{minipage}{16cm}
\begin{theorem} \label{converged}(Convergence for fdMCC, fdBMC)
If subsidy is removed at time step $d$, then within $2|E|+|V|$ time steps the evolving adoption vector will converge to a cycle of length at most 2.
\end{theorem}
\end{minipage}}\\

Next, we explain the proof of this theorem, which requires the proof of a number of intermediate lemmas.
The update rule of our network Moral-Motivation Model
is a special case of the general threshold-based update rule Goles analyzed in \cite{Dynamicsofpositiveautomatanetworks}. We follow Goles analysis closely, but by exploiting the monotonicity of our restricted case, and emphasizing a more combinatorial
description of a key function, we give significantly tighter results for our model. Further, we state an explicit bound that is linear in the size of the edge set of the network regardless of degree distribution (whereas \cite{Dynamicsofpositiveautomatanetworks} states explicit bounds only for uniform-degree networks).

In the following proof we will use the fact (from \cite{Dynamicsofpositiveautomatanetworks}) that for a set of specified integer cardinality thresholds (our model as described until now) denoted $b_i$ for $i \in V$, replacing $b_i$ with $b_i-0.5$ for all $i$ gives an identical update procedure.\footnote{E.g. replacing a threshold of 3 at node $i$ with a threshold of 2.5 changes nothing about how the adoption vector will change.} Thus, without loss of generality, we will assume all thresholds are half-integer.

\noindent \fbox{
\begin{minipage}{16cm}
\begin{lemma} \label{enterscycle}(Adoption enters a cycle \cite{Dynamicsofpositiveautomatanetworks})
Suppose the subsidy is removed at time $d$, and the adoption vector evolves according to the update rule. There exists a $c$ with the property that $x(t)=x(t+c)=x(t+2c)=...$ and that $x(t)$ is not equal to any of $x(t+1), x(t+2),...,x(t+c-1)$ for all t above some \textit{Transient Time } $T$.
\end{lemma}
\end{minipage}}\\

\noindent \textit{Proof:} The space of possible adoption vectors, $\{0,1\}^{|V|}$, is finite. Thus, at some time step an adoption vector $y$ must be repeated, and since the update process is deterministic, the behavior from that point forward will be identical to the evolution after the first occurrence of $y$, giving a cycle.$\Box$. \\

\noindent We'll define and analyze a special function $E(x(t))$. In contrast to the analysis in \cite{Dynamicsofpositiveautomatanetworks}, we describe $E(x(t))$ in terms of the update process in the network. To do this we introduce the idea of \textit{sightings}. Given an adoption vector $y \in \{0,1\}^{|V|}$, we say that node $i$ \textit{sights} each of its neighbors which is 1 according to y.\footnote{In terms of our update process:
are the number of sightings $i$ makes at $x(t)$ greater than
$b_i$? If so, then $i$ is 1 at time $t+1$.} The expression for $E(x(t))$ will include two terms related to sightings.\\

\noindent \textbf{Sightings Necessary to get $x(t)$: } \\
If $x(t)$ occurs during our updating process, we can give a bound on how many sightings must have happened
in the graph at time $t-1$: if $i$ is 1 at $t$, it must have sighted at least $b_i$ neighbors which were 1.
Thus, throughout the graph at least
\[
\sum_{i=1}^{|V|}b_ix_i(t) = \langle b,x(t) \rangle =  \text{(Sightings Necessary to get $x(t)$) }
\]
sightings must have happened at $t-1$. This counts accurately: $b_i$ is included exactly when $x_i(t)$ is 1.\\

\noindent \textbf{Sightings wasted in turning on $x(t+1)$:}\\
The adoption vector $x(t)$ produces the next state $x(t+1)$:
each node $i$ that is
1 in $x(t+1)$ saw at least $b_i$ sightings in $x(t)$, but $i$ may also have sighted
some extra neighbors at 1 beyond the $b_i$ that were required. We say \textit{these sightings were wasted in turning on $x(t+1)$}. Let $A_i$ denote the $i$th row of the adjacency matrix for $G$.\footnote{The adjacency matrix of $G$ is the $|V|\times|V|$ matrix that has $A_{ij}=1$ exactly when the edge $(i,j)$ is in $G$ and has all other entries 0.} The number of sightings that were wasted at node $i$ is
$
A_ix(t)-b_i$.
So, summing over $i$:
\begin{align*}
\sum_{i=1}^{|V|} (A_ix(t)-b_i)(x_i(t+1)) &= \langle (Ax(t)-b), x(t+1) \rangle\\
&= \text{(Sightings wasted in turning on $x(t+1)$)}
\end{align*}

\noindent This counts the correct quantity because $A_ix(t)-b_i$ is included in the sum exactly when $x_i(t+1)=1$. \\

\noindent \textbf{Defining $E(x(t))$:}
\begin{align*}
E(x(t))&:= \text{(Sightings Necessary to get $x(t)$) } - \text{ (Sightings wasted in turning on $x(t+1)$)}\\
&= \langle b,x(t)\rangle-\langle (Ax(t)-b), x(t+1) \rangle
\end{align*}

\noindent \fbox{
\begin{minipage}{16cm}
\begin{lemma} \label{decreasingunless2cycle} \cite{Dynamicsofpositiveautomatanetworks}
If the adoption vector has not entered a 2-cycle (aka, assuming $x(t)\neq x(t+2)$), then $E(x(t))$ is decreasing:
\[
E(x(t+1))+0.5 \leq  E(x(t)).
\]
\end{lemma}
\end{minipage}}\\

\noindent \textit{Proof: } The proof of this lemma is largely from \cite{Dynamicsofpositiveautomatanetworks}: we include the details for completeness and because they are critical to explaining the subsequent improvements we make.\\

\noindent We will show $E(x(t))-E(x(t+1))\geq 0.5$.
\begin{align*}
E(x(t))&-E(x(t+1))=\\
&= \langle b,x(t)\rangle-\langle (Ax(t)-b), x(t+1) \rangle - \langle b,x(t+1)\rangle +\langle (Ax(t+1)-b), x(t+2) \rangle\\
&= \langle b,x(t)\rangle -\langle (Ax(t)), x(t+1) \rangle +\langle (Ax(t+1)-b), x(t+2) \rangle\\
&= \langle b,x(t)\rangle -\langle (Ax(t+1)), x(t) \rangle +\langle (Ax(t+1)-b), x(t+2) \rangle\\
&=-\langle (Ax(t+1)-b), x(t) \rangle +\langle (Ax(t+1)-b), x(t+2) \rangle\\
&=\langle x(t+2)-x(t), (Ax(t+1)-b)\rangle
\end{align*}
In the algebra above: first we make a combined term for all sightings at $t$ by nodes which are 1 at $t+1$, next we switch the order of summation: sightings by 1s from $t+1$ at $x(t)$ are the same as sightings by 1s from $t$ at $x(t+1)$ (which is alternately true from $A$ a symmetric matrix), then simply combine terms.\\

\noindent The righthand side sums the following term over all nodes $i$:
\[
[x_i(t+2)-x_i(t)](A_ix(t+1)-b_i)
\]
If $i$ has the same state in $x(t+2)$ and $x(t)$ then the term for $i$ has value 0. \\
If $x_i(t+2)=1$ and $x_i(t)=0$: $x_i(t+2)=1$  means $A_ix(t+1)-b_i\geq0.5$\\
If $x_i(t+2)=0$ and $x_i(t)=1$: $x_i(t+2)=0$  means $A_ix(t+1)-b_i\leq-0.5$\\

\noindent These facts follow from the half-integrality of the $b_i$. In both cases where $x_i(t+2)\neq x_i(t)$, the term for node $i$ contributes at least $1/2$ to the
value of $E(x(t))-E(x(t+1))$.  Since we assumed we are not in a 2-cycle yet (aka that $x(t)\neq x(t+2)$) there must be at least one $i$ that contributes value 1/2. $\Box$ \\

\noindent We can now prove that the evolving adoption vector will converge to a cycle of length at most 2.\\

\noindent \textit{Proof (second half of Theorem \ref{converged}): }Suppose the cycle guaranteed by Lemma \ref{enterscycle} has length $c>2$ :
$x(t)=x(t+c)$ and $x(t)$ is not equal to any of $x(t+1), x(t+2),...,x(t+c-1)$. We can apply Lemma \ref{decreasingunless2cycle}:
\[
E(x(t))>E(x(t+1))>E(x(t+2))>...>E(x(t+c)).
\]
Since $x(t)=x(t+c)$, we also have that $E(x(t))=E(x(t+c))$. This gives a contradiction. Thus $c\leq 2$. $\Box$\\

We have already shown in Lemma \ref{decreasingunless2cycle} that $E(x(t))$ decreases in every time step unless the process has converged to its final 2-cycle, so bounding the range of $E(x(t))$ will give an upper bound on the transient time of the process.  This is our general strategy, but to give an improved upper bound, we use the monotonicity of our process (Lemma \ref{monoton}) to show that unless the process is already very close to a 2-cycle, the decrease in $E(x(t))$ is twice as large per time step as
specified by Lemma \ref{decreasingunless2cycle}:\\

\noindent \fbox{
\begin{minipage}{16cm}
\begin{lemma}
Unless the evolving adoption vector is within $2|V|$ time-steps of entering a 2-cycle, at least 2 nodes $i$ have $x_i(t)\neq x_i(t+2)$, so that
\[
E(x(t+1))+1\leq  E(x(t)).
\]
\end{lemma}
\end{minipage}}\\

\noindent \textit{Proof:} Suppose that $x(t+2)$ differs from $x(t)$ in only one position.  We'll show that the adoption vector enters a 2-cycle within $2|V|$ time steps. Let $N_{x(t)}$ denote the subset of $V$ corresponding to nodes that are at 1 in $x(t)$. Suppose that $x_i(t+2)=1$ and $x_i(t)=0$, and for all other $j$, $x_j(t+2)=x_j(t)$. By definition, $N_{x(t)}\subseteq N_{x(t+2)}$. We use monotonicity to reason about the set of adopters in following time steps.

Consider $N_{x(t+1)}$: it results from sightings at $N_{x(t)}$.  Since $N_{x(t)}\subseteq N_{x(t+2)}$, all sightings at $N_{x(t)}$ happen at $N_{x(t+2)}$.  Thus, $N_{x(t+3)}$ is a superset of $N_{x(t+1)}$: any node that decided to adopt based on $x(t)$ will certainly adopt based on $x(t+2)$. By the same rationale, $N_{x(t+1)}\subseteq N_{x(t+3)}$ gives that $N_{x(t+2)}\subseteq N_{x(t+4)}$. Then $N_{x(t+2)}\subseteq N_{x(t+4)}$ gives that $N_{x(t+3)}\subseteq N_{x(t+5)}$, etc.  This argument holds iteratively. To make this more clear, write the set of adopters in a z-pattern using $N_{x(t)}\Rightarrow N_{x(t+1)}$ to denote that the set of adopters at time $t+1$ results from the set of adopters at time $t$:\\
\begin{center}
\begin{tabular}{lll}
\vspace{3mm}
$N_{x(t)}$              & $\Rightarrow$ &  $N_{x(t+1)}$     \\
\vspace{3mm}
\hspace{2mm}          &$\rtoldownimplies$  &  \\
\vspace{3mm}
$N_{x(t+2)}$            & $\Rightarrow$ &   $N_{x(t+3)}$   \\
\vspace{3mm}
\hspace{2mm}          &$\rtoldownimplies$  & \\
\vspace{3mm}
$N_{x(t+4)}$            & $\Rightarrow$ &  $N_{x(t+5)}$    \\
\vspace{3mm}
\hspace{2mm}          &$\rtoldownimplies$ &  \\
\vspace{3mm}
$N_{x(t+6)}$           & $\Rightarrow$ &   $N_{x(t+7)}$    \\
\vspace{3mm}
\hspace{7mm}$\dotsdown$          &$\dotsdiag$ & \hspace{5mm}$\dotsdown$
\end{tabular}
\end{center}
\vspace{-3mm}
Now, add the subset relationships:
\begin{center}
\begin{tabular}{lll}
\vspace{3mm}
$N_{x(t)}$              & $\Rightarrow$ &  $N_{x(t+1)}$     \\
\vspace{3mm}
\hspace{2mm} $\subseteqdown$         &$\rtoldownimplies$  & $\subseteqdown$ \\
\vspace{3mm}
$N_{x(t+2)}$            & $\Rightarrow$ &   $N_{x(t+3)}$   \\
\vspace{3mm}
\hspace{2mm} $\subseteqdown$         &$\rtoldownimplies$  & $\subseteqdown$\\
\vspace{3mm}
$N_{x(t+4)}$            & $\Rightarrow$ &  $N_{x(t+5)}$    \\
\vspace{3mm}
\hspace{2mm} $\subseteqdown$         &$\rtoldownimplies$ & $\subseteqdown$ \\
\vspace{3mm}
$N_{x(t+6)}$           & $\Rightarrow$ &  $N_{x(t+7)}$ \\
\vspace{3mm}
\hspace{7mm}$\dotsdown$          &$\dotsdiag$ & \hspace{5mm}$\dotsdown$  \\
\vspace{3mm}
\end{tabular}

\end{center}

Since the update process is deterministic, if any of these $\subseteq$ relationships is not proper then the adoption vector has entered a 2-cycle. That is, if $N_{x(t+k)}=N_{x(t+k+2)}$, then it also must be the case that $N_{x(t+k+1)}=N_{x(t+k+3)}$, so that for all time steps after $t+k$ the adoption vector alternates between $x(t+k)$ and $x(t+k+1)$. The maximum number of time steps that could elapse before a $\subseteq$ relationship is forced to be non-proper is $2|V|$: since all the subsets must be proper, at least one node is in $(N_{x(t+k+2)}\setminus N_{x(t+k)})$ for all $k$. The longest path that could ever be achieved is if $N_{x(t)}$ is the empty set and $N_{x(t+2|V|)}$ is the entire set $V$.  After this many time steps it must be the case that $N_{x(t+k+2)}=N_{x(t+k)}$: the adoption vector has entered a 2-cycle within $2|V|$ time steps.

A symmetric argument (with opposite direction of $\subseteq$ relationships) establishes the case where $x_i(t+2)=0$ and $x_i(t)=1$, and for all other $j$, $x_j(t+2)=x_j(t)$.
We have established that if $x(t+2)$ differs from $x(t)$ in only one position the adoption vector will enter a 2-cycle within $2|V|$ time steps. Equivalently, if the adoption vector is more than $2|V|$ time steps from entering a 2-cycle, then it must be that if $x(t+2)$ differs from $x(t)$ in strictly more than 1 position.  Thus, evaluating the final expression from the proof of Lemma \ref{decreasingunless2cycle}:
\begin{align*}
E(x(t))&-E(x(t+1))=\langle x(t+2)-x(t), (Ax(t+1)-b)\rangle \geq 1.  \hspace{10mm} \Box.
\end{align*}





\noindent \fbox{
\begin{minipage}{16cm}
\begin{lemma}\label{range}
The range of values that $E(x(t))$ can achieve is $\leq 2|E|$.
\end{lemma}
\end{minipage}}\\



\noindent Recall the definition of $E(x(t))$:\\

$
E(x(t))=
\text{(Sightings Necessary to get $x(t)$)}$-$\text{(Sightings wasted in turning on $x(t+1)$)}\\
$

\noindent From our definitions of necessary and wasted sightings, the first term is always positive (or 0) and the second term is always negative (or 0). Letting the first term be as large as possible, and the second term be as small as possible, we obtain an upper bound on $E(x(t))$ of $\sum_{i=1}^{|V|} b_i$ (every node in $|V|$ makes the necessary sightings for it to adopt).

An obvious lower bound on $E(x(t))$ assumes the first term is $0$ and makes the second term as large in magnitude
as possible (as many wasted sightings as possible at every node):
\[
E(x(t))\geq -\sum_{i=1}^{|V|} (\text{deg}(i)-b_j)= -\sum_{i=1}^{|V|} (\text{deg}(i))+\sum_{i=1}^{|V|} b_i.
\]
Thus the range of $E(x(t))$ is at most
\[
\sum_{i=1}^{|V|} (\text{deg}(i))=2|E|. \hspace{10mm} \Box.
\]

\noindent \textit{Proof:} (Convergence time claimed in Theorem \ref{converged})
We start from the end of the subsidy period. The range of $E(x(t))$ is at most $2|E|$ from Lemma \ref{range}. At most $2|V|$ steps can decrease $E(x(t))$ by only 1/2.  Every other step must have size at least 1.  Thus, there are at most $(2|V|+ (2|E|-|V|)/1)=2|E|+|V|$ time steps after the subsidy is removed before the adoption vector enters a 2-cycle. $\Box.$

\section{Appendix: Results on Hardness of Approximation:}

\noindent \begin{tabular}{|l|l|l|}\hline
& \textbf{temporary (temp)} & \textbf{fixed-duration (fd)} \\ \hline
\textbf{Min-Cost} &    &   \\
\textbf{Complete}            &  $\Omega(ln (|V|) )$)        & $\Omega(ln (|V|) )$    \\
\textbf{Conversion}            &        &     (provided that $d\geq 3$)      \\ \hline
\textbf{Budgeted}&     &  \\
\textbf{Maximum}           &  $<1-\frac{1}{e}$ $\approx 0.632$  &    $<1-\frac{1}{e}$ $\approx 0.632$          \\
\textbf{Conversion}           &    &         (provided that $d\geq 3$)  \\ \hline
\end{tabular}
\vspace{4mm}

For both cases we reduce to problems with hardness guarantees due to Feige \cite{Feige:1998:TLN:285055.285059}. Both reductions will rely on constructing instances with intransigent nodes (that have $\alpha_i>1$).

Consider the \textit{Set Cover Problem} defined as follows. Let $S$ denote a set of elements $\{1,2,...,n\}$. Let $F$ denote a group of subsets of $S$ which we will denote $J_1,J_2,...,J_{|F|}$. The objective is to specify the smallest set of indices $I$ so that
\[
\cup_{i\in I}J_i=S.\\
\]

\noindent \fbox{
  \begin{minipage}{16cm}
\begin{theorem}
The Set Cover Problem can be reduced in polynomial time to an instance of Min-cost Complete Conversion (for any subsidy duration of at least 3 time steps).
\end{theorem}
\end{minipage}}

\begin{proof}(Reducing to Min-cost Complete Conversion (for arbitrary $d\geq 3$)): \\
Given an instance of the Set Cover Problem, construct an instance of Min-cost Complete Conversion as follows.  Construct a graph with three sets of nodes: for each subset $J_i$ introduce a \textit{subset node} called $n_{J_i}$, and for each element $q$ introduce an \textit{element node} called $n_q$. If element $q$ is in subset $J_i$ then add the edge $(n_q, n_{J_i})$ to the graph. Finally, each element node $n_q$ has a unique dummy copy called $n'_q$ and the edge $(n_q, n'_q)$ is in the graph. See the figure below.

\begin{figure}[ht!]\label{fig:moralutil}
  \centering
  \fbox{
  \begin{minipage}{12cm}
    \includegraphics[trim = 30mm 93mm 40mm 90mm, clip,
width=10cm]{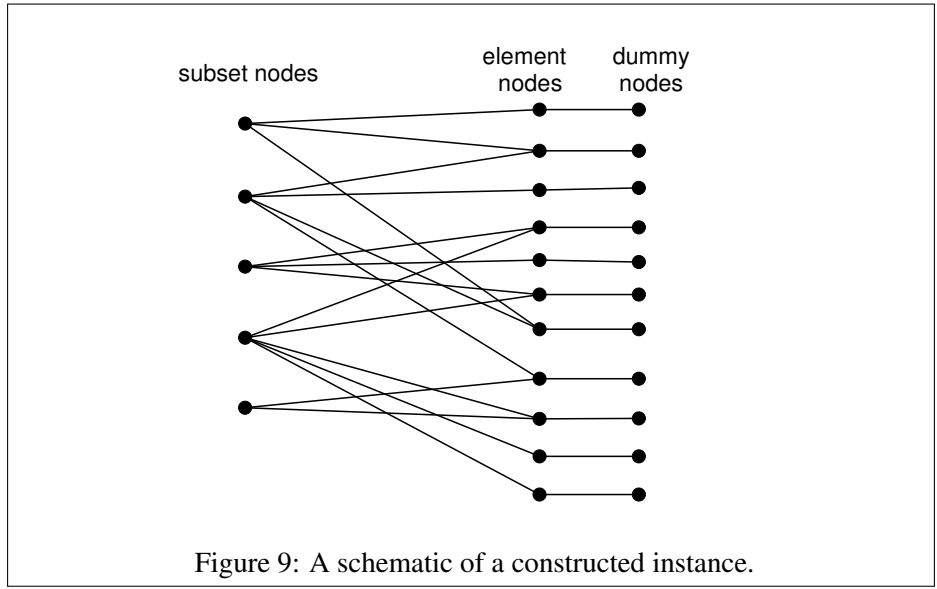} \caption{A schematic of a constructed instance.}
\end{minipage}}
\end{figure}
For every element node, let $\alpha_{n_q}= 1/(|F|+1)$.
For every subset node, let $\alpha_{n_{J_i}}=1+\epsilon$ for some small $\epsilon>0$, and for every dummy node let $\alpha_{n'_q}=1$.

Notice that subset nodes will adopt only when they are directly subsidized. Thus, all element nodes that will ever adopt (for any reason) will do so by the end of the third time step at latest. If any single neighbor of an element node adopts, the element node will also adopt. If an element node and its dummy node are both adopters when subsidy is removed, they will both be permanent adopters.

Given a black-box to solve Min-cost Complete Conversion: we get the smallest set to subsidize that converts all element and dummy nodes permanently after a subsidy of duration $d$.\footnote{This is the most permanent adoption that can be achieved since all subset nodes have $\alpha_{n_{J_i}}>1$}  Call this current optimal subsidy set $O$. We explain how to find a subsidy set consisting only of subset nodes which has size $|O|$ and converts all element and dummy nodes permanently (within the same duration of subsidy).

Suppose that $O$ contains a dummy node $n'_q$. Then the updated set $(O\cup n_q)\backslash n'_q$ has size at most $|O|$ and converts all element and dummy nodes permanently after a subsidy of duration $d$. Notice that if $|(O\cup n_q)\backslash n'_q|<|O|$, this would contradict the optimality of $O$. Repeat until all dummy nodes have been removed from $O$.

Suppose that the updated $O$ contains an element node $n_q$. Since $O$ is the smallest set that permanently converts all element nodes, some subset node that is a neighbor of $n_q$ must not be in $O$ (otherwise removing $n_q$ from $O$ gives a smaller subsidy set that converts all element and dummy nodes within duration $d$). Let $n_{J_k}$ denote such an unsubsidized neighbor.  The alternate subsidy set $(O\cup n_{J_k})\backslash n_q$ has size $|O|$ and converts all element and dummy nodes permanently within duration $d$. Repeat until all element nodes have been removed from $O$.

Now we have a set of $|O|$ subset nodes that permanently converts all element and dummy nodes.  By our construction, this means that the set of subset indices corresponding to our subsidy set has resulting union equal to the entire set of elements $S$ (since an element node adopts precisely when a subset node that is its neighbor is subsidized).  Thus, we have a solution of size $|O|$ for the set cover instance.

For the original set cover instance, suppose that a strictly smaller solution for the set cover instance exists.  Then subsidizing the corresponding set of subset nodes in our constructed Min-cost Complete Conversion instance would give a subsidy set that permanently converts all element and dummy nodes for any subsidy of length at least 3 time steps (all element nodes would be converted in the second time step, all dummy nodes in the third time step), but has size $<|O|$. This contradicts the optimality of $O$ for the constructed instance.

Thus, a black-box for Min-cost complete conversion gives a polynomial-time algorithm to solve the set cover instance exactly. Thus Min-cost Complete Conversion (for arbitrary $d\geq 3$) inherits $\Omega(ln (|V|) )$-hardness from the Set Cover Problem (this hardness holds unless NP has slightly superpolynomial time algorithms, see \cite{Feige:1998:TLN:285055.285059} for details).
\end{proof}

Consider the \textit{Budgeted Maximum-coverage Problem} defined as follows. Let $S$ denote a set of elements $\{1,2,...,n\}$. Let $F$ denote a group of subsets of $S$ which we will denote $J_1,J_2,...,J_{|F|}$. Given a budget $k$, the objective is to specify a set of $k$ indices $I$ so that the cardinality of the following union is maximized:
\[
\cup_{i\in I}J_i.\\
\]

\noindent \fbox{
  \begin{minipage}{16cm}
\begin{theorem}
The Budgeted Maximum-coverage Problem can be reduced in polynomial time to an instance of Budgeted Maximum Conversion (for any subsidy duration of at least 3 time steps).
\end{theorem}
\end{minipage}}

\begin{proof}
The graph constructed is identical to the previous case. We make the additional observation that for subsidy duration of at least 3 the permanent adoption status of an element node and its dummy are identical (either both are adopters or neither adopts).

Given a black-box to solve Budgeted Maximum Conversion: we get a size-$k$ subsidy set that converts the maximum possible number of (element, dummy)-pairs of nodes permanently after a subsidy of duration $d$. Call the number of pairs converted $h$. As in the case above we can produce a subsidy set of the same size which permanently converts an identical set of (element, dummy)-pairs but is composed only of subset nodes.

Now we have a set of $k$ subset nodes that permanently converts $h$ (element, dummy)-pairs.  By our construction, this means that the set of subset indices corresponding to our subsidy set has resulting union of cardinality $h$ (since an element node adopts precisely when a subset node that is its neighbor is subsidized).  Thus, we have a solution of size $k$ with union containing $h$ members of $S$ for the budgeted maximum-coverage instance.

For the original budgeted maximum-coverage instance, suppose that some other solution of size at most $k$ has union of cardinality strictly greater than $h$.  Then subsidizing the corresponding set of subset nodes in our constructed budgeted maximum-conversion instance would give a subsidy set of size at most $k$ which permanently converts strictly more than $h$ (element, dummy)-pairs for any subsidy of length at least 3 time steps (the element nodes of each such pair would be converted in the second time step, the dummy nodes of each such pair in the third time step). This contradicts the optimality of $O$ for the constructed instance.

Thus, a black-box for budgeted maximum conversion gives a polynomial-time algorithm to solve the budgeted maximum coverage instance exactly. Thus budgeted maximum conversion (for arbitrary $d\geq 3$) inherits $(1-1/e)$-hardness from the Budgeted Maximum Coverage Problem (this hardness holds unless P=NP, see \cite{Feige:1998:TLN:285055.285059} for details).

\end{proof}
\vspace{-8mm}
\section{Appendix: IP formulation for additional problem variants}
\subsection{temporary Budgeted Maximum Conversion (tempBMC)}
Decision Variables:
\vspace{-2mm}
\begin{itemize}
\item \textit{(Subsidy Variables)} for $i\in V$:
\[
 y_i = \begin{cases}
        1  & \text{ if node $i$ is subsidized from time step 0 to $|V|$}\\
        0  & \text{ otherwise}
        \end{cases}
\]
\vspace{-5mm}
\item \textit{(Adoption Variables)} for $i\in V, t\in \{0,1,2,..., 2|V|\}$:
\[
 x_{it} = \begin{cases}
        1  & \text{ if node $i$ adopts at time $t$}\\
        0  & \text{ otherwise}
        \end{cases}
\]
\vspace{-5mm}
\end{itemize}
The number of nodes chosen for subsidy is at most the given budget $k$. During the period of the subsidy, a node $i$ is allowed to be an adopter only if either it is subsidized, or if at least $b_i$ of its neighbors are adopters during the previous time step. As explained for tempMCC, assuming subsidy is maintained for $|V|$ time steps perfectly emulates the conditions we defined for temporary subsidy. After the period of subsidy ends, a node $i$ is allowed to be an adopter only if at least $b_i$ of its neighbors are adopters during the previous time step. As explained for tempMCC, imposing these conditions for $|V|$ time steps ensures that the process will have converged to a stable adoption vector: counting the adopters in the final time step gives the number of long term adopters (precisely the quantity we seek to maximize).  For given budget, $k$, the following IP computes the largest set of long term adopters that can be achieved by the strategy of subsidizing $k$ nodes until growth in adoption has stopped, and then removing all subsidy.

\begin{align*}
\text{maximize} \sum_{i} x_{i,2|V|} \hspace{5mm} & \\
 \text{subject to} \hspace{5mm} &\\
  \sum_{i\in V} y_i &\leq k\\
   x_{it}           &\leq y_i+\frac{1}{b_i}\sum_{j\in \delta(i)} x_{j,t-1}\hspace{5mm}        \text{ for } t \in \{0,1,2,..., |V|\},  i \in V.\\
   x_{it}           &\leq \frac{1}{b_i}\sum_{j\in \delta(i)} x_{j,t-1}\hspace{5mm}                 \text{ for } t \in \{ |V|+1,|V|+2, ..., 2|V|\}, i \in V.
\end{align*}
\vspace{-5mm}

\subsection{fixed-duration Min-cost Complete Conversion (fdMCC)}

Decision Variables:
\vspace{-2mm}
\begin{itemize}
\item \textit{(Subsidy Variables)} for $i\in V$:
\[
 y_i = \begin{cases}
        1  & \text{ if node $i$ is subsidized from time step 0 to $d$}\\
        0  & \text{ otherwise}
        \end{cases}
\]
\vspace{-4mm}
\item \textit{(Adoption Variables)} for $i\in V, t\in \{0,1,2,..., (d+2|E|+|V|)\}$:
\[
 x_{it} = \begin{cases}
        1  & \text{ if node $i$ adopts at time $t$}\\
        0  & \text{ otherwise}
        \end{cases}
\]
\vspace{-4mm}

\item \textit{(Subsidy Size)} a dummy variable $q$ describes the size of the subsidized set.
\end{itemize}
Minimize $q$ subject to the following constraints.  The number of nodes chosen for subsidy is at most $q$. During the period of the subsidy of $d$ time steps, a node $i$ is allowed to be an adopter only if either it is subsidized, or if at least $b_i$ of its neighbors are adopters during the previous time step.  After the period of subsidy ends, a node $i$ is allowed to be an adopter only if at least $b_i$ of its neighbors are adopters during the previous time step. From Theorem \ref{converged}, imposing these conditions for $2|E|+|V|$ time steps ensures that the process will have converged to a 2-cycle.  Forcing the adoption variables for the final time 2 time steps ensures we only consider solutions which result in $100\%$ adoption. For given subsidy duration $d$, the following IP computes the smallest subsidy set which, when subsidized until growth in adoption stops, permanently converts the entire network.
\vspace{-2mm}
\begin{align*}
\text{minimize } q \hspace{5mm} & \\
 \text{subject to} \hspace{5mm} &\\
  \sum_{i\in V} y_i &\leq q\\
   x_{it}           &\leq y_i+\frac{1}{b_i}\sum_{j\in \delta(i)} x_{j,t-1}\hspace{5mm}        \text{ for } t \in \{0,1,2,..., d\},  i \in V.\\
   x_{it}           &\leq \frac{1}{b_i}\sum_{j\in \delta(i)} x_{j,t-1}\hspace{5mm}                 \text{ for } t \in \{ d+1,d+2, ..., (d+2|E|+|V|)\}, i \in V.\\
   x_{i,(d+2|E|+|V|-1)}& \geq 1 \hspace{5mm}                 \text{ for }  i \in V.\\
   x_{i,(d+2|E|+|V|)}& \geq 1 \hspace{5mm}                 \text{ for }  i \in V.
\end{align*}
\vspace{-5mm}

\subsection{fixed-duration Budgeted Maximum Conversion (fdBMC)}

Decision Variables:
\vspace{-2mm}
\begin{itemize}
\item \textit{(Subsidy Variables)} for $i\in V$:
\[
 y_i = \begin{cases}
        1  & \text{ if node $i$ is subsidized from time step 0 to $d$}\\
        0  & \text{ otherwise}
        \end{cases}
\]
\vspace{-2mm}
\item \textit{(Adoption Variables)} for $i\in V, t\in \{0,1,2,..., (d+2|E|+|V|)\}$:
\[
 x_{it} = \begin{cases}
        1  & \text{ if node $i$ adopts at time $t$}\\
        0  & \text{ otherwise}
        \end{cases}
\]
\end{itemize}
\vspace{-2mm}
The number of nodes chosen for subsidy is at most the given budget $k$. During the period of the subsidy of $d$ time steps, a node $i$ is allowed to be an adopter only if either it is subsidized, or if at least $b_i$ of its neighbors are adopters during the previous time step.  After the period of subsidy ends, a node $i$ is allowed to be an adopter only if at least $b_i$ of its neighbors are adopters during the previous time step. From Theorem \ref{converged}, imposing these conditions for $2|E|+|V|$ time steps ensures that the process will have converged to a 2-cycle. Thus, averaging the adoption variables for the final time 2 time steps gives the rate of long term average adoption (the quantity we wish to maximize). For given subsidy duration $d$, and budget $k$, the following IP computes the subsidy set which gives the highest long term average adoption rate.
\vspace{-2mm}
\begin{align*}
\text{maximize } \frac{1}{2}\Big( \sum_{i \in V}&x_{i,(d+2|E|+|V|)}+ \sum_{i \in V}x_{i,(d+2|E|+|V|-1)}\Big)\hspace{5mm}  \\
 \text{subject to} \hspace{5mm} &\\
  \sum_{i\in V} y_i &\leq k\\
   x_{it}           &\leq y_i+\frac{1}{b_i}\sum_{j\in \delta(i)} x_{j,t-1}\hspace{5mm}        \text{ for } t \in \{0,1,2,..., d\},  i \in V.\\
   x_{it}           &\leq \frac{1}{b_i}\sum_{j\in \delta(i)} x_{j,t-1}\hspace{5mm}                 \text{ for } t \in \{ d+1,d+2, ..., (d+2|E|+|V|)\}, i \in V.\\
\end{align*}


\end{document}